\documentclass[a4paper,11pt]{article}
\usepackage{bm,mathrsfs,amsmath,amsthm,amssymb,amscd,longtable,array,graphicx,bm}
\newcommand{\nc}{\newcommand}
\nc{\rnc}{\renewcommand}
\nc{\nn}{\nonumber}
\nc{\der}{{\partial}}
\rnc{\Im}{{\rm{Im}\,}}
\rnc{\Re}{{\rm{Re}\,}}
\nc{\db}{\displaybreak[0]\\}
\nc{\bra}{\langle}
\nc{\ket}{\rangle}
\nc{\bs}{\boldsymbol}

\newtheorem{theorem}{Theorem}[section]
\newtheorem{lemma}[theorem]{Lemma}

\newtheorem{proposition}[theorem]{Proposition}

\theoremstyle{definition}

\numberwithin{equation}{section}

\numberwithin{equation}{section}

\begin{document}%
%
\title{
Yang-Baxter algebra, higher rank partition functions \\
and $K$-theoretic Gysin map
for partial flag bundles
}

\author{
Kohei Motegi \thanks{E-mail: kmoteg0@kaiyodai.ac.jp}
\\\\
{\it Faculty of Marine Technology, Tokyo University of Marine Science and Technology,}\\
 {\it Etchujima 2-1-6, Koto-Ku, Tokyo, 135-8533, Japan} \\
\\\\
\\
}

\date{\today}

\maketitle

\begin{abstract}
We investigate the $K$-theoretic Gysin map for type $A$ partial flag bundles
from the viewpoint of integrability.
We introduce several types of partition functions for one version of
$q=0$ degeneration of $U_q(\widehat{sl_n})$ vertex models on rectangular grids
which differ by boundary conditions and sizes,
and can be viewed as Grothendieck classes of the Grothendieck group
of a nonsingular variety and partial flag bundles.
By deriving multiple commutation relations for the $q=0$ $U_q(\widehat{sl_n})$ Yang-Baxter algebra and combining with the
description of the $K$-theoretic Gysin map for partial flag bundles
using symmetrizing operators,
we show that the $K$-theoretic Gysin map of the first type of
partition functions on a rectangular grid is given by the
second type whose boundary conditions on one side are reversed from the first type.
This generalizes the author's previous result
from Grassmann bundles to partial flag bundles.
We also discuss the inhomogenous version of
the partition functions and applications
to the $K$-theoretic Gysin map.

\end{abstract}

\section{Introduction}

Integrable systems, which originated from physics
\cite{Bethe,LW,Baxter,KBI}, have deep influence not only in physics
but also mathematics.
One of the most celebrated examples is the birth of quantum groups
\cite{Dr,J} from the quantum inverse scattering method \cite{KBI,FST,KRS}.
In recent years, exploring connections between
algebraic geometry, integrable systems, representation theory and combinatorics
is an active research area in mathematics and mathematical physics.
See 
\cite{GRTV,MS,MS2,RTV,AObethe,Okounkov,MO,AO,GRV,Konno1,Konno2,IIM,PSZ,KPSZ,KZ,SO,KirillovSigma,GK,WZ,BS,BFHTW,ZJ,Yeliussizov,Iwao,Iwaoskew,Iwaorefinedskew,MotegiGrassmann,GMSZ} for examples for various topics.

In our previous work \cite{MotegiGrassmann},
by observing and using the similarity between
the multiple commutation relations of the Yang-Baxter algebra
of a certain five-vertex model and
the formula for $K$-theoretic pushforward (which is
also called as $K$-theoretic Gysin map) for Grassmann bundles,
we showed that the pushforward of the Grothendieck classes of Grassmann bundles
represented by one type of partition functions
of the five-vertex model on a rectangular grid
are given by the Grothendieck classes of a nonsingular variety
represented by another type of partition functions
on a rectangular grid which differ by boundary conditions.
The results obtained have
intersections with the formulas derived by Buch \cite{Buch1}
(cohomological version is given in
\cite{Pragacz,FP} for example), where Grassmannian Grothendieck polynomials
appear. See \cite{LS,FK,Mc,IN,Buch2,Buch3,Buch4,BuchMihalcea,Lenart} for examples for seminal works on Grothendieck polynomials and studies on related geometric aspects.

In this paper, we extend our previous study from
Grassmann bundles to type $A$ partial flag bundles.
We study the $K$-theoretic pushforward of Grothendieck classes
of partial flag bundles from the viewpoint of quantum integrability.
The integrable models which we use are one version of
the $q=0$ degeneration of the $U_q(\widehat{sl_n})$ vertex models.
From the Yang-Baxter algebra associated to the vertex models,
we can derive the following multiple commutation relations
\begin{align}
&\displaystyle
\prod_{j=1}^{q_1} D_m(u_j) \prod_{j=q_{1}+1}^{q_2} B_{m-1}(u_j)
\cdots \prod_{j=q_{m-1}+1}^{q_m} B_1(u_j) \prod_{j=q_m+1}^{n} B_0(u_j)  \nonumber \\
=&\sum_{\overline{w} \in S_n/S_{q_1} \times S_{q_2-q_1} \times \cdots \times S_{n-q_m}} w \cdot \Bigg[ 
\frac{1}
{\prod_{k=1}^m \prod_{q_{k-1}<i \leq q_k} \prod_{q_{k}<j \leq n}(1-u_j/u_i)}
\nonumber \\
&\times 
\prod_{j=q_m+1}^{n} B_0(u_j)
\prod_{j=q_{m-1}+1}^{q_m} B_1(u_j)
\cdots
\prod_{j=q_{1}+1}^{q_2} B_{m-1}(u_j)
\prod_{j=1}^{q_1} D_m(u_j)
\Bigg]. \label{introductionhigherrankmultiplecommutationrelations}
\end{align}
Details of the notations etc will be explained later.
This type of multiple commutation relations
of the Yang-Baxter algebra has similarity with the 
following $K$-theoretic Gysin formula for
partial flag bundles using symmetrizing operators
\begin{align}
&\pi_{q_1,\dots,q_m *} (f(1-u_1^{-1},\dots,1-u_n^{-1})) \nonumber \\
=&\sum_{\overline{w} \in S_n/S_{q_1} \times S_{q_2-q_1} \times \cdots \times S_{n-q_m}} w \cdot \Bigg[ 
\frac{f(1-u_1^{-1},\dots,1-u_n^{-1})}
{\prod_{k=1}^m \prod_{q_{k-1}<i \leq q_k} \prod_{q_{k}<j \leq n}(1-u_j/u_i)}
\Bigg], \label{introductionkgysinformulauvariables}
\end{align}
which the details of the notations in the equation above will
also be explained later.

In this paper, we introduce two
classes of partition functions of higher rank vertex models
on a rectangular grid which differ by boundary conditions
and can be regarded as Grothendieck classes of the Grothendieck group
of partial flag bundles and a nonsingular variety respectively.
Exploring connections between partition functions of higher rank models
(which are also called as colored lattice models) and
probability theory, number theory, combinatorics and other areas of
mathematical physics is a hot research topic.
See \cite{Ko,Iz,Ku,Ts,PRS,Ros,Wheeler,Motegi,HK1,Bogo,ShigechiUchiyama,BBF,BMN,Lascoux,Mcnamara,KS,Korff,BW,BWZ,WZ2,Borodin,BP1,Takeyama,vDE,BBB,BBBGduality,BBBGdem,BorodinWheeler,Zhong,FM,Motegihigherrank,ABW,BBBG}
as well as papers \cite{MS,MS2,GK,WZ,BS,BFHTW} mentioned before
for examples on these topics as well as
seminal and previous studies on partition functions.
Using
\eqref{introductionhigherrankmultiplecommutationrelations} and
\eqref{introductionkgysinformulauvariables},
we show two types of higher rank partition functions introduced
are directly connected by the $K$-theoretic Gysin map,
which generalizes the formula for Grassmann bundles in
the author's previous work.
From the viewpoint of integrability, it is well-known that one can
introduce inhomogeneous parameters into partition functions.
We also discuss applications of the inhomogeneous version to
the $K$-theoretic Gysin map.

This paper is organized as follows.
In the next section, we introduce the $R$-matrix and monodromy matrices
which we use in this paper.
In section three, we introduce three types of higher rank
partition functions. The first two types are partition functions
on a rectangular grid which differ by boundary conditions
on one side.
We also introduce another type of partition functions
on a larger rectangular grid with simpler boundary conditions than the first two types,
and show this type is equivalent to the second
type of the higher rank partition functions
which represents Grothendieck classes of a nonsingular variety.
In section four,
we introduce the higher rank Yang-Baxter algebra and derive the
multiple commutation relations used in this paper. We also discuss
symmetries of the partition functions introduced in section three.
In section five, combining the
multiple commutation relations for the Yang-Baxter algebra and the
description of $K$-theoretic Gysin map for partial flag bundles
using symmetrizing operators,
we show the $K$-theoretic pushforward of the partition functions on a rectangular grid corresponding to the Grothendieck classes of partial flag bundles
is given by the ones
corresponding to the Grothendieck classes of a nonsingular variety.
In the last section,
we introduce the inhomogenous version of the partition functions and
discuss applications to the $K$-theoretic Gysin map.

\section{Higher rank vertex models}

In this section, we introduce the $R$-matrix,
monodromy matrix and its matrix elements with respect to the auxiliary space
which are used to construct several types of partition functions and
the Yang-Baxter algebra associated to the vertex models used in this paper.

Let $W$ be an $(m+1)$-dimensional complex vector space
and denote its standard basis as $\{ | 0 \rangle, |1 \rangle, \dots, |m \rangle  \}$.
We denote the dual of $|k \rangle$ as $\langle k|$ ($k=0,1,\dots,m$).
The dual vector space is denoted as $W^*$, which is spanned by
 $\{ \langle 0|, \langle 1|, \dots, \langle m|  \}$.
In this paper, we use the bra-ket notation without taking complex conjugation.
For example, the orthogonality of standard basis
is expressed as
$\langle k| \ell \rangle=\delta_{k \ell}$, $k,\ell=0,1,\dots,m$.

We frequently adopt conventional notations
used in quantum integrable models and quantum groups.
We refer to \cite{MotegiGrassmann} for example for
a detailed account of the notations.
We use subscripts to distinguish vector spaces.
For example, we denote the tensor product of
two $m$-dimensional vector spaces as $W_i \otimes W_j$.
We can take $\{ |k \rangle_i \otimes |\ell \rangle_j \ | \ k,\ell=0,1,\dots,m
 \}$ as a basis of $W_i \otimes W_j$.

The $R$-matrix $R_{ij}(u,w)$ acting on $W_i \otimes W_j$
is defined by acting on this basis as
\begin{align}
R_{ij}(u,w)|k \rangle_i \otimes |k \rangle_j
&=|k \rangle_i \otimes |k \rangle_j, \ \ \ k=0,1,\dots,m, \label{rdefone} \\
R_{ij}(u,w)|k \rangle_i \otimes |\ell \rangle_j
&=w/u|\ell \rangle_i \otimes |k \rangle_j, \ \ \ 0 \leq k < \ell \leq m, \label{rdeftwo} \\
R_{ij}(u,w)|k \rangle_i \otimes |\ell \rangle_j
&=|\ell \rangle_i \otimes |k \rangle_j
+(1-w/u)|k \rangle_i \otimes |\ell \rangle_j,
\ \ \ m \geq k > \ell \geq 0,
\label{rdefthree}
\end{align}
where $u$ and $w$ are complex numbers, which are usually called as spectral parameters.
The subscripts $i$ and $j$ of $R_{ij}(u,w)$ are used to indicate the spaces
the $R$-matrix act.

One of the important properties of this $R$-matrix is the following one
which is
often called as the ice-rule:
${}_i \langle \delta_1| \otimes {}_j \langle \delta_2|
R(u,w)| \epsilon_1  \rangle_i \otimes | \epsilon_2 \rangle_j=0$
if 
$\epsilon_1+\epsilon_2 \neq \delta_1+\delta_2$.

\begin{figure}[ht]
\includegraphics[width=11cm]{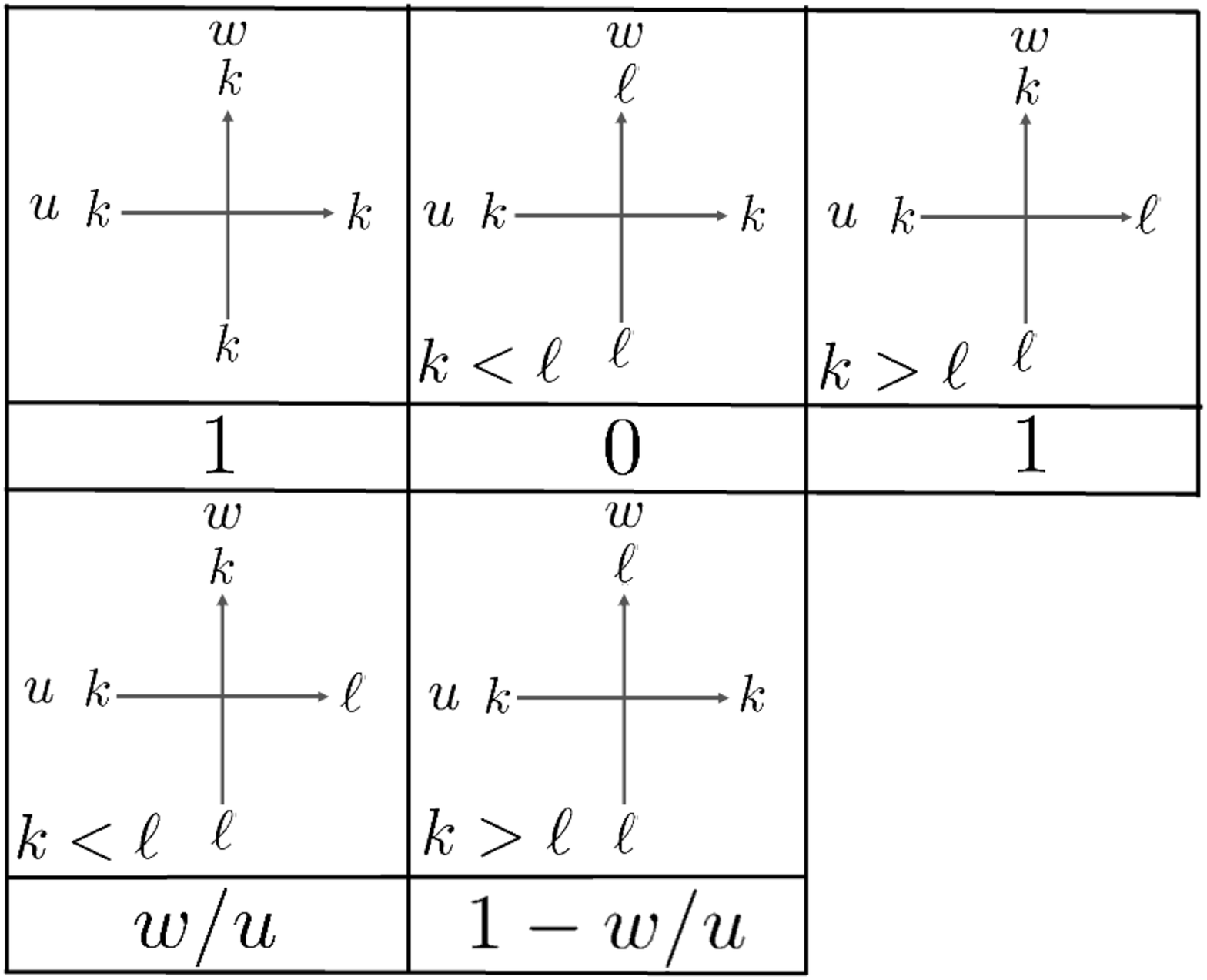}
\caption{A graphical description of the $R$-matrix $R_{ij}(u,w)$
which is a $q=0$ degeneration of the $U_q(\widehat{sl_{m+1}})$ $R$-matrix
\eqref{rdefone}, \eqref{rdeftwo}, \eqref{rdefthree}.
We display the  elements of the $R$-matrix.
The $R$-matrix is represented as two crossing arrows.
The left and the up arrow represents the space $W_i$
and $W_j$, respectively. 
The top left figure represents
${}_i \langle k | \otimes {}_j \langle k| R(u,w)|k \rangle_i \otimes |k \rangle_j=1$, $0 \le k \le m$.
The top middle figure represents
${}_i \langle k | \otimes {}_j \langle \ell| R(u,w)|k \rangle_i \otimes |\ell \rangle_j=0$, $0 \le k < \ell \le m$.
The top right figure represents
${}_i \langle \ell | \otimes {}_j \langle k| R(u,w)|k \rangle_i \otimes |\ell \rangle_j=1$, $m \ge k > \ell \ge 0$.
The bottom left figure represents
${}_i \langle \ell | \otimes {}_j \langle k| R(u,w)|k \rangle_i \otimes |\ell \rangle_j=w/u$, $0 \le k < \ell \le m$.
The bottom right figure represents
${}_i \langle k | \otimes {}_j \langle \ell| R(u,w)|k \rangle_i \otimes |\ell \rangle_j=1-w/u$, $m \ge k > \ell \ge 0$.
All the other matrix elements which are not displayed in the
figure above are identically zero.
}
\label{picturermatrix}
\end{figure}

The $R$-matrix is a
higher rank generalization of the one used for example in \cite{MS,MS2,GK,WZ}
and can be regarded
as a $q=0$ limit of the $U_q(\widehat{sl_{m+1}})$ $R$-matrix
\cite{Dr,J}, a different limit from the ones which have direct connections with
the crystal basis theory. The $U_q(\widehat{sl_{m+1}})$ version
can be found in \cite{BFHTW,BorodinWheeler,Zhong} for example.
Let us write down the action of the $U_q(\widehat{sl_{m+1}})$ $R$-matrix $R_{ij}^q(u,w)$ on basis vectors for completeness.
\begin{align}
R_{ij}^q(u,w)|k \rangle_i \otimes |k \rangle_j
&=(u-qw)|k \rangle_i \otimes |k \rangle_j, \ \ \ k=0,1,\dots,m,  \\
R_{ij}^q(u,w)|k \rangle_i \otimes |\ell \rangle_j
&=q(u-w)|k \rangle_i \otimes |\ell \rangle_j+
(1-q)w|\ell \rangle_i \otimes |k \rangle_j, \ \ \ 0 \leq k < \ell \leq m,  \\
R_{ij}^q(u,w)|k \rangle_i \otimes |\ell \rangle_j
&=(1-q)u|\ell \rangle_i \otimes |k \rangle_j
+(u-w)|k \rangle_i \otimes |\ell \rangle_j,
\ \ \ m \geq k > \ell \geq 0.
\end{align}
The $R$-matrix $R_{ij}(u,w)$ which we use in this paper is obtained
from $R_{ij}^q(u,w)$ as $R_{ij}(u,w)=u^{-1} R_{ij}^q(u,w)|_{q=0}$.
In recent years, the $R$-matrix $(u-qw)^{-1}R_{ij}^q(u,w)$ is sometimes
referred to as stochastic $R$-matrix.
The $R$-matrices have origins in statistical physics.
See Figure \ref{picturermatrix} for a graphical
description of the $R$-matrix which we use in this paper.
We construct monodromy matrices and partition functions
from this $R$-matrix.
Also note here that we denote
$U_q(\widehat{sl_{m+1}})$ $R$-matrix
instead of $U_q(\widehat{sl_n})$ $R$-matrix
so as to be consistent with the notations in the algebraic geometry or algebraic topology side,
which we adopt the notations by Nakagawa-Naruse 
\cite{NN1,NN3}.

Traditionally, we also use the notation $R_{ij}(u,w)$
for operators acting on a larger space than $W_i \otimes W_j$.
More generally, for an integer $p \geq 2$, we define
$R_{ij}(u,w)$ $(1 \le i < j \le p)$
as an operator acting on $W_1 \otimes W_2 \otimes \cdots \otimes
W_p$. The action is defined as follows:
$R_{ij}(u,w)$ acts on the tensor product of $W_i$ and $W_j$ as
\eqref{rdefone}, \eqref{rdeftwo}, \eqref{rdefthree},
and acts as identity
on all the other vector spaces $W_k$, $k \neq i,j$.

\begin{figure}[ht]
\includegraphics[width=12cm]{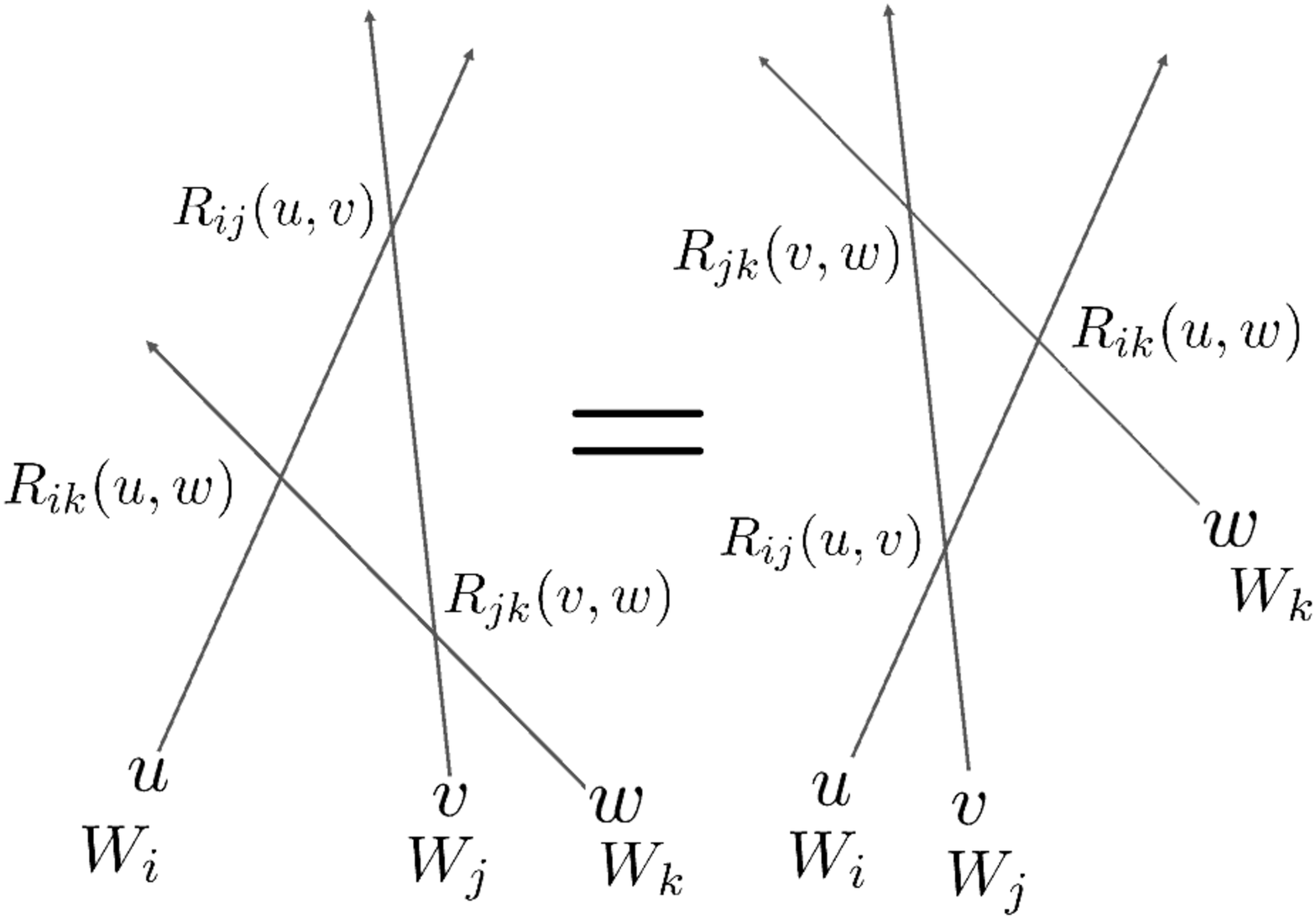}
\caption{A graphical description of the Yang-Baxter relation $R_{ij}(u,v)R_{ik}(u,w)R_{jk}(v,w)
=R_{jk}(v,w)R_{ik}(u,w)R_{ij}(u,v) \in \mathrm{End}(W_i \otimes W_j \otimes W_k)$ \eqref{yangbaxter}.
The left and right figure represents $R_{ij}(u,v)R_{ik}(u,w)R_{jk}(v,w)$
and $R_{jk}(v,w)R_{ik}(u,w)R_{ij}(u,v)$ respectively.
}
\label{pictureyangbaxter}
\end{figure}

The $R$-matrix $R_{ij}(u,w)$
satisifies the Yang-Baxter relation
(Figure \ref{pictureyangbaxter})
\begin{align}
R_{ij}(u,v)R_{ik}(u,w)R_{jk}(v,w)
=R_{jk}(v,w)R_{ik}(u,w)R_{ij}(u,v). \label{yangbaxter}
\end{align}
We regard \eqref{yangbaxter}
as a relation on $W_i \otimes W_j \otimes W_k$ and
view $R_{ij}(u,v)$, $R_{ik}(u,w)$ and $R_{jk}(v,w)$ as operators
acting on $W_i \otimes W_j \otimes W_k$,
which acts nontrivially on $W_i \otimes W_j$,
$W_i \otimes W_k$ and $W_j \otimes W_k$ respectively,
and acts as identity on $W_k$, $W_j$ and $W_i$ respectively.

More generally, we view \eqref{yangbaxter}
as a relation in a larger vector space
$W_1 \otimes \cdots \otimes W_p$ for an integer $p \geq 3$
and regard that
$R_{ij}(u,v)$ acts nontrivially on $W_i$ and $W_j$,
$R_{ik}(u,w)$ nontrivially on $W_i$ and $W_k$,
$R_{jk}(v,w)$ nontrivially on $W_j$ and $W_k$,
and each of them acts as identity on all the other spaces.

From the $R$-matrix,
we construct the monodromy matrix $T_a(u)$
\begin{align}
T_{a}(u)&=R_{a, p}(u,1) \cdots R_{a 1}(u,1),
\label{monodromy}
\end{align}
which are operators acting on $W_a \otimes W_1 \otimes \cdots \otimes W_{p}$.
The $R$-matrix $R_{a, j}(u,1)$ $(1 \le j \le p)$ acts nontrivially on $W_a$ and $W_j$ and as identity on the remaining spaces.
All the $R$-matrices act nontrivially on $W_a$,
and this space is traditionally called as the auxiliary space.
We use alphabets, typlically $a$
to label this special space and denote the vector space as $W_a$.
On the other hand, the other spaces $W_1, \dots, W_p$ are called as quantum spaces.

Since the monodromy matrix $T_a(u)$ 
acts on $W_a \otimes W_1 \otimes \cdots \otimes W_{p}$ whose dimension
is $(m+1)^{p+1}$, this monodromy matrix 
can be regarded as
a $(m+1)^{p+1} \times (m+1)^{p+1}$ matrix.

We now consider
the matrix elements of the monodromy matrix
with respect the auxiliary space $W_a$
\begin{align}
T_{ij}(u):={}_a \langle j|T_a(u)|i \rangle_a, \ \ \ i,j=0,\dots,m.
\end{align}
Each of the elements $T_{ij}(u)$ act on $W_1 \otimes \cdots \otimes W_p$
and can be regarded as $(m+1)^p \times (m+1)^p$ matrices.

\begin{figure}[ht]
\includegraphics[width=10cm]{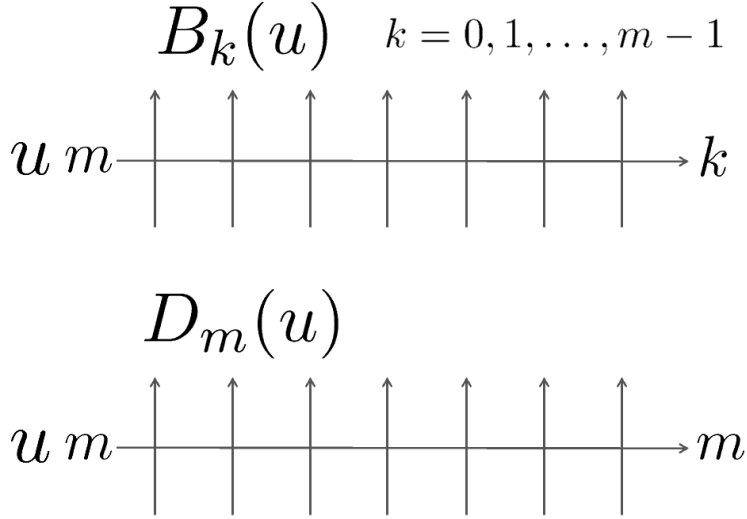}
\caption{The operators
$B_k(u)=T_{mk}(u)={}_a \langle k|T_a(u)|m \rangle_a$ ($k=0,1,\dots,m-1$) \eqref{bkop} (top)
and $D_m(u)=T_{mm}(u)={}_a \langle m|T_a(u)|m \rangle_a$ \eqref{dmop} (bottom).
}
\label{picturebandd}
\end{figure}

In this paper, we use the following ones
\begin{align}
B_k(u):&=T_{mk}(u), \ \ \ k=0,1,\dots,m-1, \label{bkop} \\
D_m(u):&=T_{mm}(u). \label{dmop}
\end{align}
See Figure \ref{picturebandd} for graphical descriptions of the operators
$B_k(u)$ ($k=0,1,\dots,m-1$) and $D_m(u)$.

\section{Partition functions}
We introduce three types of partition functions
in this section.
Let us first introduce notations for the basis of
the space $W_1 \otimes W_2 \otimes \cdots \otimes W_p$
and its dual $W_1^* \otimes W_2^* \otimes \cdots \otimes W_p^*$.
For an ordered set of integers $I=\{i_1,i_2,\dots,i_p \}$ satisfying
$0 \leq i_1,i_2,\dots,i_p \leq m$, we define $|I \rangle$ and $\langle I|$
as
\begin{align}
|I \rangle&=|i_1 \rangle_{1} \otimes |i_2 \rangle_{2} \otimes \cdots \otimes
|i_p \rangle_p \in W_1 \otimes W_2 \otimes \cdots \otimes W_p, \\
\langle I|&={}_1 \langle i_1| \otimes {}_2 \langle i_2| \otimes \cdots \otimes
{}_p \langle i_p| \in W_1^* \otimes W_2^* \otimes \cdots \otimes W_p^*.
\end{align}
$\{ |I \rangle \ | \ 0 \leq i_1,i_2,\dots,i_p \leq m \}$
and $\{ \langle I | \ | \ 0 \leq i_1,i_2,\dots,i_p \leq m \}$
forms a standard basis of $W_1 \otimes W_2 \otimes \cdots \otimes W_p$
and $W_1^* \otimes W_2^* \otimes \cdots \otimes W_p^*$ respectively.

\begin{figure}[ht]
\includegraphics[width=12cm]{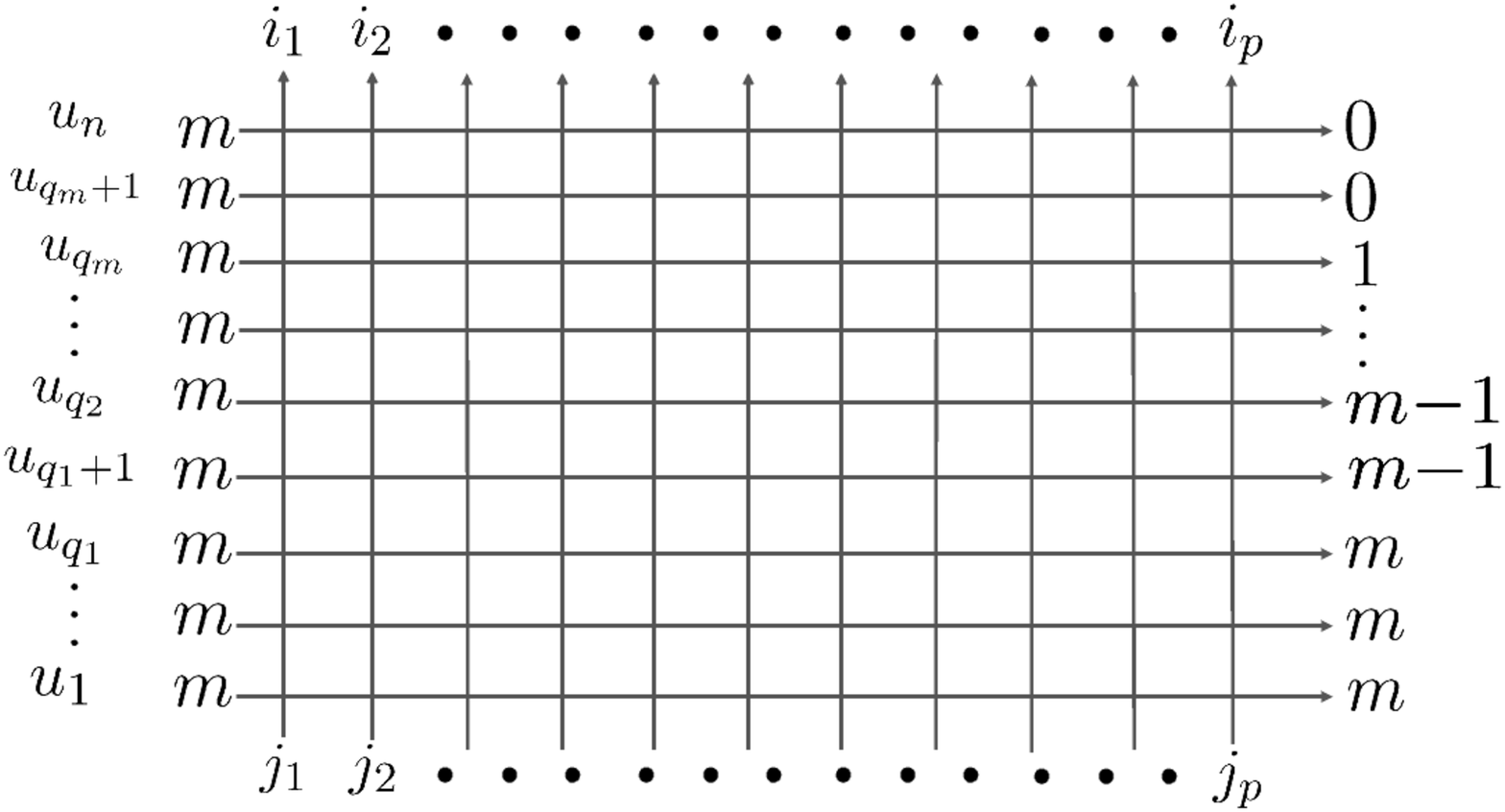}
\caption{The partition functions
$F_{IJ}(u_1,\dots,u_n)$
\eqref{beforepushforwardpartitionfunction}.
The top sequence of integers corresponds to $\langle I|$.
The first $n-q_m$ rows correspond to $B_0(u_{q_m+1}),\dots,B_0(u_n)$.
The next $q_m-q_{m-1}$ rows correspond to
$B_1(u_{q_{m-1}+1}),\dots,B_1(u_{q_m})$, and so on.
The bottom sequence of integers corresponds to
$|J \rangle$.
}
\label{picturebeforepushforward}
\end{figure}

\begin{figure}[ht]
\includegraphics[width=12cm]{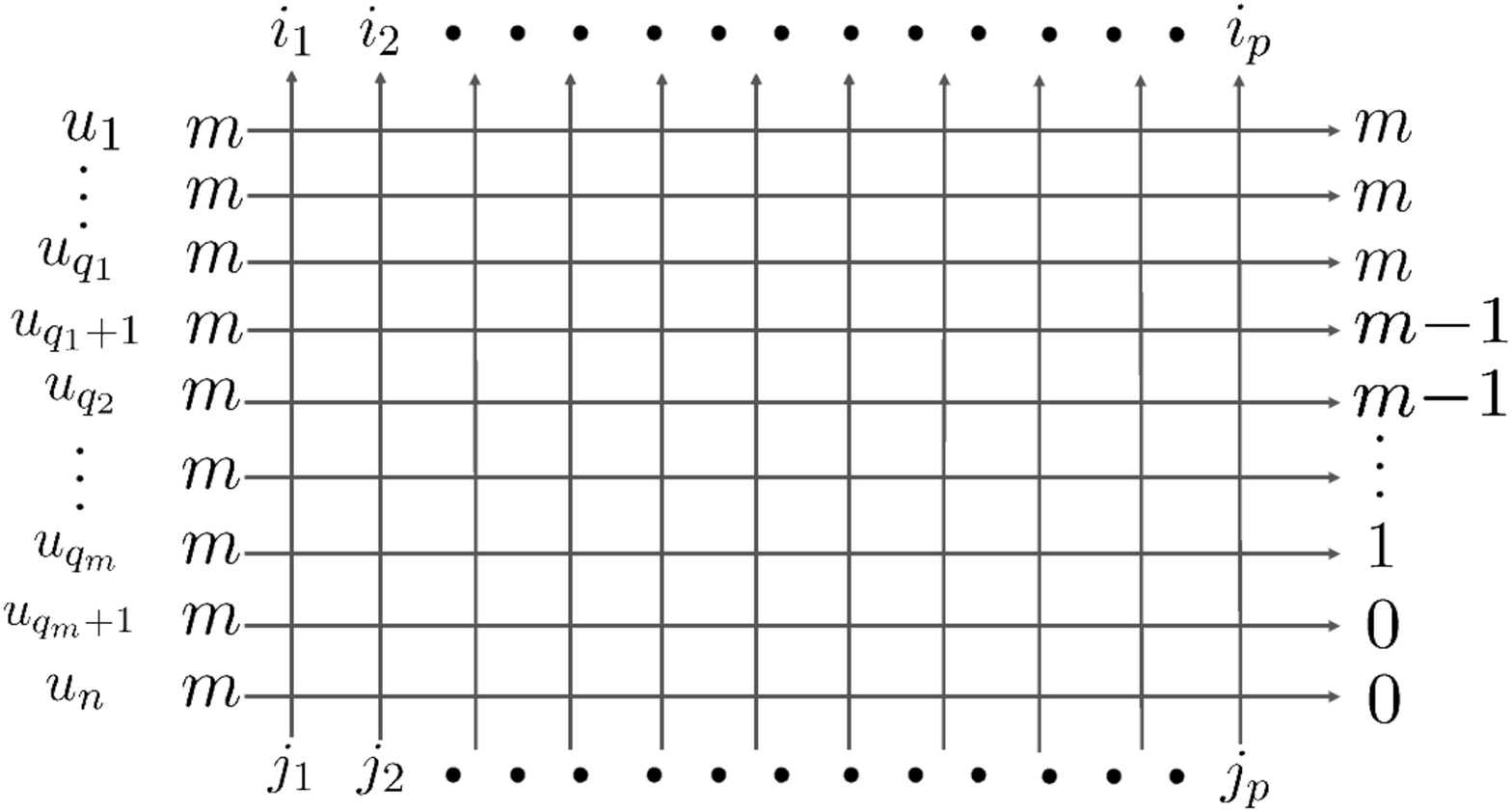}
\caption{The partition functions
$G_{IJ}(u_1,\dots,u_n)$
\eqref{afterpushforwardpartitionfunction}.
Note that since the ordering of operators is reversed from
$F_{IJ}(u_1,\dots,u_n)$, the boundary condition
on the right is reversed from
Figure \ref{picturebeforepushforward}.
We also reversed the ordering of spectral parameters.
However, the ordering of the spectral parameters do not matter
since $G_{IJ}(u_1,\dots,u_n)$ are symmetric with respect to
$u_1,\dots,u_n$. This symmetry can be understood for example by showing that
$G_{IJ}(u_1,\dots,u_n)$
are equivalent to the third type of partition functions
$H_{IJ}(u_1,\dots,u_n)$ constructed only from
$B_0$-operators. Then the symmetry follows from
the commutativity of the $B_0$-operators.
}
\label{pictureafterpushforward}
\end{figure}

We also introduce a set of integers $q_1,q_2,\dots,q_m$ satisfying
$q_0:=0<q_1<q_2<\cdots< q_m<q_{m+1}:=n$.

Using the operators $B_k(u)$ ($k=0,1,\dots,m-1$) \eqref{bkop},
$D_m(u)$ \eqref{dmop}
and (dual) basis vectors $\langle I|$, $|J \rangle$,
we introduce the first type of partition functions $F_{IJ}(u_1,\dots,u_n)$
as (see Figure \ref{picturebeforepushforward} for a graphical description)
\begin{align}
&F_{IJ}(u_1,\dots,u_n) \nonumber \\
=&
\langle I|
\prod_{j=q_m+1}^{n} B_0(u_j)
\prod_{j=q_{m-1}+1}^{q_m} B_1(u_j)
\cdots
\prod_{j=q_{1}+1}^{q_2} B_{m-1}(u_j)
\prod_{j=1}^{q_1} D_m(u_j) |J \rangle.
\label{beforepushforwardpartitionfunction}
\end{align}

\begin{figure}[ht]
\includegraphics[width=12cm]{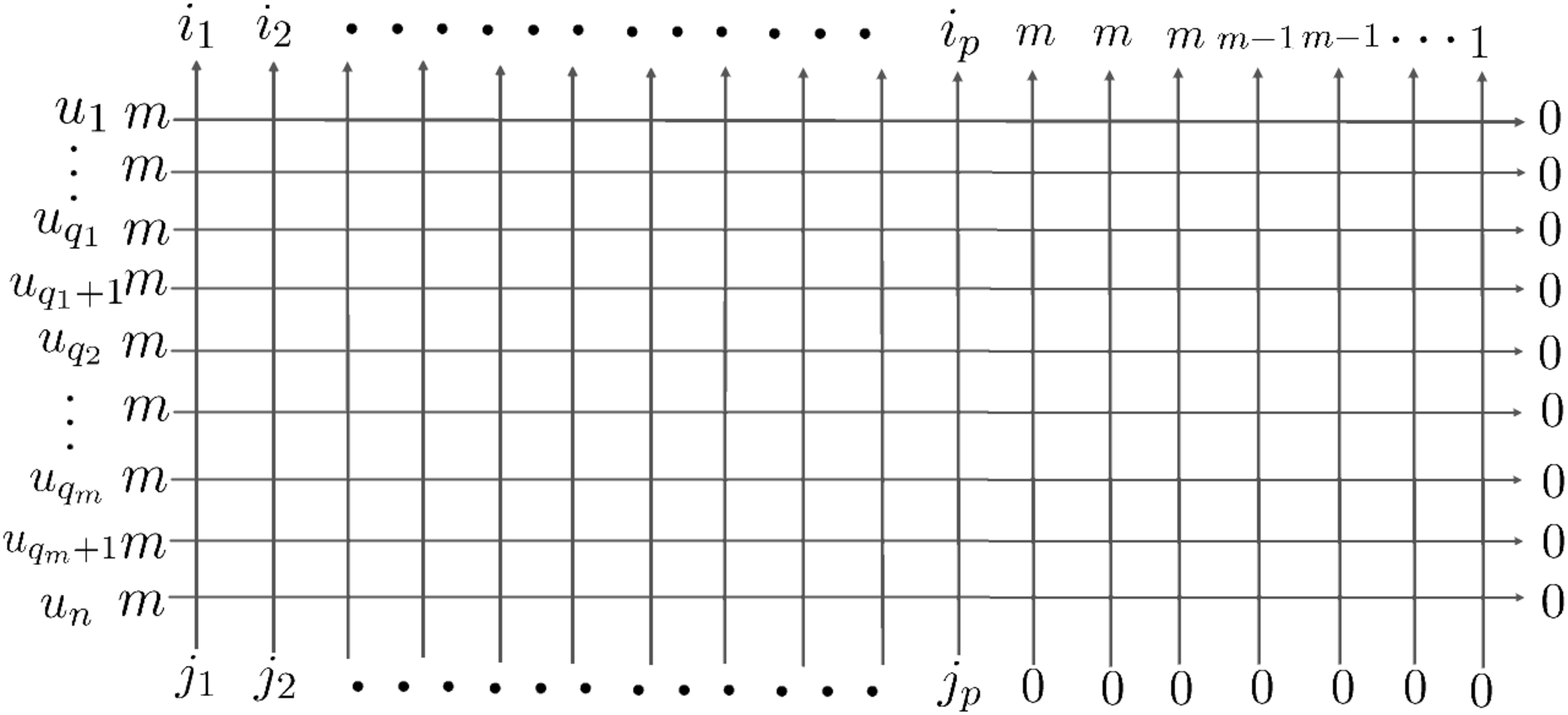}
\caption{The partition functions
$H_{IJ}(u_1,\dots,u_n)$
\eqref{largerandequivalent}. Note the rectangular grid is wider than
the ones for $F_{IJ}(u_1,\dots,u_n)$ and $G_{IJ}(u_1,\dots,u_n)$.
$q_m$ columns are added,
and  integers are also added in top and bottom of those columns
as in the figure.
Note also that the integers in the right boundary are all 0,
which means that the partition functions are constructed only
using the $B_0$-operators. Since the $B_0$-operators are
commutative, this implies that $H_{IJ}(u_1,\dots,u_n)$
are symmetric with respect to $u_1,\dots,u_n$.
}
\label{pictureequivalentafterpushforward}
\end{figure}

We next introduce the second type of partition functions 
$G_{IJ}(u_1,\dots,u_n)$ as

\begin{align}
&G_{IJ}(u_1,\dots,u_n) \nonumber \\
=&
\langle I|
\prod_{j=1}^{q_1} D_m(u_j) \prod_{j=q_{1}+1}^{q_2} B_{m-1}(u_j)
\cdots \prod_{j=q_{m-1}+1}^{q_m} B_1(u_j) \prod_{j=q_m+1}^{n} B_0(u_j)
|J \rangle.
\label{afterpushforwardpartitionfunction}
\end{align}
Note the order of operators to define $G_{IJ}(u_1,\dots,u_n)$ in
\eqref{afterpushforwardpartitionfunction}
is reversed from the one for $F_{IJ}(u_1,\dots,u_n)$
\eqref{beforepushforwardpartitionfunction}.
See Figure \ref{pictureafterpushforward} for a graphical description.

\begin{figure}[ht]
\includegraphics[width=12cm]{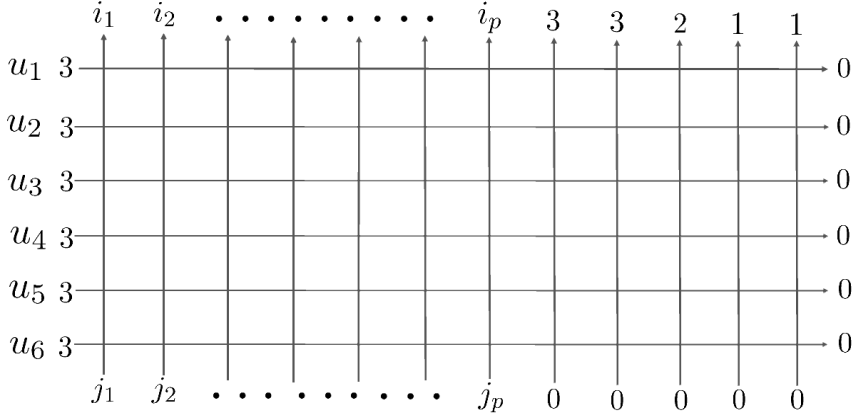}
\caption{The partition functions $H_{IJ}(u_1,\dots,u_n)$
corresponding to
$m=3$, $n=6$, $q_1=2$, $q_2=3$, $q_3=5$.
}
\label{pictureexoneequivalentafterpushforward}
\end{figure}

\begin{figure}[ht]
\includegraphics[width=12cm]{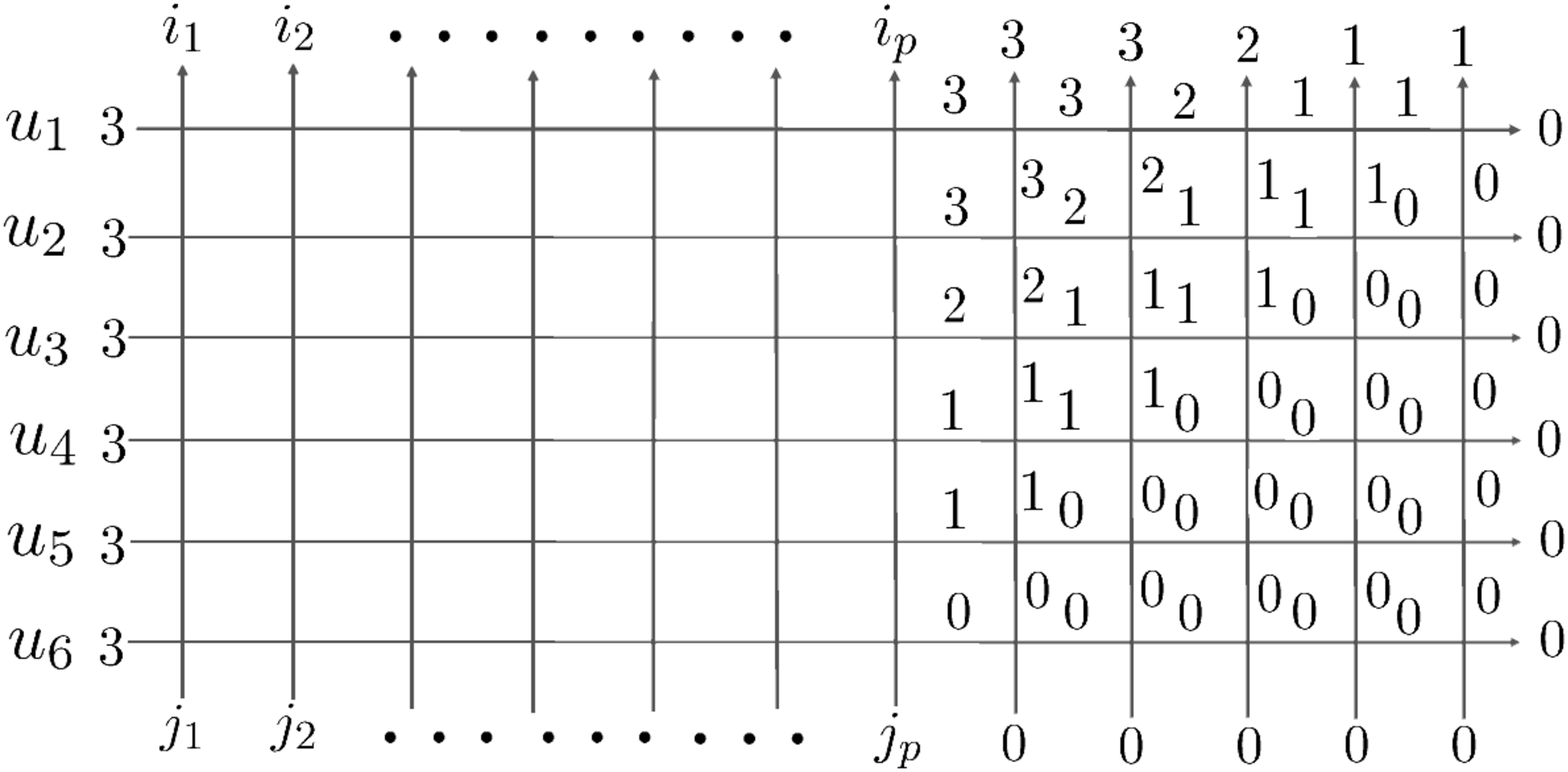}
\caption{The  partition functions
$H_{IJ}(u_1,\dots,u_n)$
corresponding to
$m=3$, $n=6$, $q_1=2$, $q_2=3$, $q_3=5$.
Using Figure \ref{picturermatrix}, one observes that only
one configuration is allowed in the rightmost $q_m$ columns.
The remaining part is nothing but the  partition functions
$G_{IJ}(u_1,\dots,u_n)$.
}
\label{pictureextwoequivalentafterpushforward}
\end{figure}

We also introduce the third type of partition functions
which is equivalent to the second type $G_{IJ}(u_1,\dots,u_n)$.
The third type uses only $B_0$-operators and
acts on a larger quantum space.
Let us define $B_0^{(p,q_1,q_2,\dots,q_m)}(u)$ as the $B_0$-operator
acting on $W_1 \otimes W_2 \otimes \cdots \otimes W_{p+q_m}$
\begin{align}
B_0^{(p,q_1,q_2,\dots,q_m)}(u)
={}_a \langle 0|
R_{a, p+ q_m}(u,1) \cdots R_{a2}(u,1) R_{a1}(u,1)|m \rangle_a.
\end{align}
We also extend the (dual) basis vectors $|I \rangle \in W_1 \otimes W_2 \otimes
\cdots \otimes W_p$ and $\langle J| \in
W_1^* \otimes W_2^* \otimes
\cdots \otimes W_p^*
$ to (dual) vectors in $W_1 \otimes W_2 \otimes
\cdots \otimes W_{p+ q_m}$ and $W_1^* \otimes W_2^* \otimes
\cdots \otimes W_{p+ q_m}^*$.
For ordered sets of integers $I=\{i_1,i_2,\dots,i_p \}$ ($0 \le i_1,i_2,\dots,i_p \le m$) and $J=\{j_1,j_2,\dots,j_p \}$ ($0 \le j_1,j_2,\dots,j_p \le m$),
we define $\langle \widetilde{I}|
\in W_1^* \otimes W_2^* \otimes
\cdots \otimes W_{p+ q_m}^*
$ and $| \widetilde{J} \rangle \in
W_1 \otimes W_2 \otimes
\cdots \otimes W_{p+ q_m}
$
as
\begin{align}
\langle \widetilde{I}|
&=
\langle I| \otimes
\langle m^{q_1},(m-1)^{q_2-q_1},\dots,2^{q_{m-1}-q_{m-2}} ,1^{q_m-q_{m-1}}|,
\\
|\widetilde{J} \rangle&=
|J \rangle \otimes |0^{ q_m} \rangle ,
\end{align}
where
\begin{align}
&| 0^{ q_m} \rangle
=|0 \rangle_{p+1} \otimes \cdots \otimes |0 \rangle_{p+q_m},
\\
&
\langle m^{q_1},(m-1)^{q_2-q_1},\dots,2^{q_{m-1}-q_{m-2}} ,1^{q_m-q_{m-1}}|
\nonumber \\
=&( {}_{p+1} \langle m| \otimes \cdots \otimes
{}_{p+q_1} \langle m|)
\otimes ({}_{p+q_1+1} \langle m-1 | \otimes \cdots \otimes
{}_{p+q_2} \langle m-1 |)
\nonumber \\
&\otimes \cdots \otimes
({}_{p+q_{m-2}+1} \langle 2 | \otimes \cdots \otimes
{}_{p+q_{m-1}} \langle 2 |) \otimes
({}_{p+q_{m-1}+1} \langle 1 | \otimes \cdots \otimes
{}_{p+q_m} \langle 1 |).
\end{align}

Using $B_0^{(p,q_1,q_2,\dots,q_m)}(u)$, $\langle \widetilde{I}|$ and $|\widetilde{J} \rangle$,
we introduce the third type of partition functions
$H_{IJ}(u_1,\dots,u_n)$ as (Figure \ref{pictureequivalentafterpushforward})
\begin{align}
H_{IJ}(u_1,\dots,u_n)=\langle \widetilde{I}|
B_0^{(p,q_1,q_2,\dots,q_m)}(u_1)
B_0^{(p,q_1,q_2,\dots,q_m)}(u_2)
\cdots
B_0^{(p,q_1,q_2,\dots,q_m)}(u_n)
|\widetilde{J} \rangle.
\label{largerandequivalent}
\end{align}

We can show that the partition functions
$G_{IJ}(u_1,\dots,u_n)$
\eqref{afterpushforwardpartitionfunction}
are the same with $H_{IJ}(u_1,\dots,u_n)$ \eqref{largerandequivalent},
i.e. they are represented by the same symmetric functions.
\begin{lemma} \label{equivalencerelation}
The following relation holds:
\begin{align}
G_{IJ}(u_1,\dots,u_n)=H_{IJ}(u_1,\dots,u_n)
. \label{equivalencetwotypes}
\end{align}
\end{lemma}
\begin{proof}
The equality \eqref{equivalencetwotypes} can be shown graphically
in the same way with Appendix C in \cite{MotegiGrassmann}.
Let us illustrate this by an example. Set
$m=3$, $n=6$, $q_1=2$, $q_2=3$, $q_3=5$,
which Figure
\ref{pictureexoneequivalentafterpushforward} corresponds to this example.
We first look the $R$-matrix at the upper right corner.
The output of the $R$-matrix of the auxiliary space and quantum space $W_{p+q_m}$ is 0 and 1 respectively.
Looking the top middle and top right figures in Figure \ref{picturermatrix},
one notes that the input of auxiliary space and  quantum space $W_{p+q_m}$
must be fixed to 1 and  0 respectively.
This means that the integers around the vertex at the upper right corner are
fixed as depicted in Figure
\ref{pictureextwoequivalentafterpushforward}.
Next, we look the $R$-matrix which is left to the one which we have just investigated. The outputs of the auxiliary space and quantum space $W_{p+q_m-1}$
are both 1.
Looking the top left figure in Figure \ref{picturermatrix}, we note
the inputs of the auxiliary space and quantum space $W_{p+q_m-1}$
must be fixed both to 1. The integers around the vertex corresponding to this $R$-matrix
are all fixed to 1. Continuing this observation,
we find only one configuration is allowed in the rightmost $q_m$ columns as
Figure
\ref{pictureextwoequivalentafterpushforward}.
The remaining part is nothing but 
$G_{IJ}(u_1,\dots,u_n)$, hence we find $H_{IJ}(u_1,\dots,u_n)$
is equal to $G_{IJ}(u_1,\dots,u_n)$ multiplied by all the $R$-matrix elements
which come from the frozen rightmost $q_m$ columns.
Since all the $R$-matrix elements which appear in those columns are 1,
we conclude $H_{IJ}(u_1,\dots,u_n)=G_{IJ}(u_1,\dots,u_n)$.
\end{proof}

\section{Yang-Baxter algebra, symmetries
and multiple commutation relations}

In this section, we show the key
multiple commutation relations of the Yang-Baxter algebra.
Let us first recall the basic commutation relations
for the operators $T_{ij}(u)$ which are matrix elements
of the monodromy matrix with respect to the auxiliary space.
For readers unfamiliar with this argument,
we refer to \cite{MotegiGrassmann} for example for a detailed account
where the meaning of the conventions used in quantum integrable models
and quantum groups are explained (on which space the operators act, for example).

From the Yang-Baxter relation \eqref{yangbaxter},
the intertwining relation between monodromy matrices follows
\begin{align}
R_{ab}(u_1,u_2) T_{a}(u_1) T_{b}(u_2)
=T_{b}(u_2) T_{a}(u_1)R_{ab}(u_1,u_2).
\label{intertwiningrelation}
\end{align}
The intertwining relation \eqref{intertwiningrelation},
which is often called as the $RTT$ relation,
can be regarded as an equality between matrices,
and equalities between the matrix elements of both hand sides are actually commutation relations among
the matrix elements $T_{ij}(u)$ of the monodromy matrix.
The commutation relations we need in this paper are the following:
\begin{align}
D_m(u_1)B_j(u_2)&=\frac{1}{1-u_2/u_1}B_j(u_2)D_m(u_1)
+\frac{1}{1-u_1/u_2}B_j(u_1)D_m(u_2), \ \ \ j=0,1,\dots,m-1, \label{rtt1} \\
B_j(u_1)B_k(u_2)&=\frac{1}{1-u_2/u_1}B_k(u_2)B_j(u_1)
+\frac{1}{1-u_1/u_2}B_k(u_1)B_j(u_2), \ \ \ 0 \leq k < j \leq m-1, \label{rtt2} \\
D_m(u_1)B_j(u_2)&=D_m(u_2)B_j(u_1), \ \ \ j=0,1,\dots,m-1, \label{rtt3} \\
B_j(u_1)B_k(u_2)&=B_j(u_2)B_k(u_1), \ \ \ 0 \leq k \leq j \leq m-1, \label{rtt4} \\
D_m(u_1)D_m(u_2)&=D_m(u_2)D_m(u_1). \label{rtt5}
\end{align}
From the $j=k$ case of \eqref{rtt4} and \eqref{rtt5},
the following symmetries hold for
the partition functions
$F_{IJ}(u_1,\dots,u_n)$ \eqref{beforepushforwardpartitionfunction}.
\begin{lemma} \label{partialsymmetries}
$F_{IJ}(u_1,\dots,u_n)$ are symmetric with respect to $u_{q_{j-1}+1},\dots,u_{q_j}$ for each j=1,\dots,m+1. Here we set $q_0:=0$, $q_{m+1}:=n$.
\end{lemma}
The $j=k$ case of \eqref{rtt4} and \eqref{rtt5}
also imply that the same symmetries hold for
$G_{IJ}(u_1,\dots,u_n)$ \eqref{afterpushforwardpartitionfunction}.
\begin{lemma}
$G_{IJ}(u_1,\dots,u_n)$ are symmetric with respect to $u_{q_{j-1}+1},\dots,u_{q_j}$ for each j=1,\dots,m+1.
\end{lemma}
We note a higher symmetry holds for $G_{IJ}(u_1,\dots,u_n)$.
These partition functions are actually fully symmetric with respect to
$u_1,\dots,u_n$.
To see this, we first note
from the $j=k=0$ case of \eqref{rtt4}
that the third type of partition functions
$H_{IJ}(u_1,\dots,u_n)$ constructed using only $B_0$-operators
\eqref{largerandequivalent} are fully symmetric.
\begin{lemma} \label{hfullsymmetry}
$H_{IJ}(u_1,\dots,u_n)$ are symmetric with respect to
$u_1,\dots,u_n$.
\end{lemma}
Combining Lemma \ref{hfullsymmetry} with
Lemma \ref{equivalencerelation},
we note $G_{IJ}(u_1,\dots,u_n)$ are fully symmetric.
\begin{lemma}
$G_{IJ}(u_1,\dots,u_n)$ are symmetric with respect to
$u_1,\dots,u_n$.
\end{lemma}

Let us now show the multiple commutation relations
from the basic commutation relations
\eqref{rtt1}, \eqref{rtt2}, \eqref{rtt3}, \eqref{rtt4} and
\eqref{rtt5}.
The simplest case $m=1$ was previously used
in \cite{MotegiGrassmann,GSFNR} to derive a skew generalization
of the identities for (factorial) Schur/Grothendieck polynomials
by Feh\'er-N\'emethi-Rim\'anyi and Guo-Sun \cite{FNR,GS}.

\begin{proposition}
The following multiple commutation relations hold:
\begin{align}
&\displaystyle
\prod_{j=1}^{q_1} D_m(u_j) \prod_{j=q_{1}+1}^{q_2} B_{m-1}(u_j)
\cdots \prod_{j=q_{m-1}+1}^{q_m} B_1(u_j) \prod_{j=q_m+1}^{n} B_0(u_j)  \nonumber \\
=&\sum_{\overline{w} \in S_n/S_{q_1} \times S_{q_2-q_1} \times \cdots \times S_{n-q_m}} w \cdot \Bigg[ 
\frac{1}
{\prod_{k=1}^m \prod_{q_{k-1}<i \leq q_k} \prod_{q_{k}<j \leq n}(1-u_j/u_i)}
\nonumber \\
&\times 
\prod_{j=q_m+1}^{n} B_0(u_j)
\prod_{j=q_{m-1}+1}^{q_m} B_1(u_j)
\cdots
\prod_{j=q_{1}+1}^{q_2} B_{m-1}(u_j)
\prod_{j=1}^{q_1} D_m(u_j)
\Bigg]. \label{higherrankmultiplecommutationrelations}
\end{align}
Here, $w$ in the right hand side of \eqref{higherrankmultiplecommutationrelations}
acts as permutation
on the the spectral parameters $(u_1,u_2,\dots,u_n)$,
and the sum is over all representatives of elements of
the fixed subgroup
$S_n/S_{q_1} \times S_{q_2-q_1} \times \cdots \times S_{n-q_m}$
of the symmetric group $S_n$.
\end{proposition}
\begin{proof}
We illustrate the simplest nontrivial case $m=2$ beyond $m=1$,
which can be applied to the cases $m>2$ as well.
We can apply the argument in \cite{ShigechiUchiyama} for this type of
higher rank multiple commutation relations.
Let us write down the commutation relations more explicitly.
Each $\overline{w} \in S_n/S_{q_1} \times S_{q_2-q_1} \times S_{n-q_2}$
can be represented as a tuple $(S^{(0)}, S^{(1)}, S^{(2)})$
where $S^{(0)}$, $S^{(1)}$, $S^{(2)}$ are unordered sets of integers such that
\begin{align}
|S^{(0)}|=q_1, \ \ \ |S^{(1)}|=q_2-q_1, \ \ \ |S^{(2)}|=n-q_2, \ \ \ S^{(0)} \cup S^{(1)} \cup S^{(2)}=\{1,2,\dots,n \}. \label{condition}
\end{align}
Then we note the case $m=2$ of
\eqref{higherrankmultiplecommutationrelations}
can be written as
\begin{align}
&\displaystyle
\prod_{j=1}^{q_1} D_2(u_j) \prod_{j=q_1+1}^{q_2} B_1(u_j) \prod_{j=q_2+1}^n B_0(u_j)  \nonumber \\
=&\sum_{(S^{(0)}, S^{(1)}, S^{(2)})} 
\prod_{\substack{i \in S^{(0)} \\ j \in S^{(1)}}} \frac{1}{1-u_j/u_i}
\prod_{\substack{i \in S^{(0)} \\ j \in S^{(2)}}} \frac{1}{1-u_j/u_i}
\prod_{\substack{i \in S^{(1)} \\ j \in S^{(2)}}} \frac{1}{1-u_j/u_i}
\nonumber \\
&\times \prod_{j \in S^{(2)}} B_0(u_j)
\prod_{j \in S^{(1)}} B_1(u_j) \prod_{j \in S^{(0)}} D_2(u_j)
. \label{higherrankmultiplecommutationrelationsm=2}
\end{align}
Here, the sum in the right hand side of 
\eqref{higherrankmultiplecommutationrelationsm=2} is
over all tuples $(S^{(0)}, S^{(1)}, S^{(2)})$ of unordered sets
of integers $S^{(0)}$, $S^{(1)}$, $S^{(2)}$
satisfying \eqref{condition}.

Let us show \eqref{higherrankmultiplecommutationrelationsm=2}.
First, starting from the left hand side,
we reverse the order of the types of the operators
using \eqref{rtt1}, \eqref{rtt2}, \eqref{rtt4} and \eqref{rtt5}
and move all $D_2$-operators to the right of $B_0$- and $B_1$-operators,
and move all $B_0$-operators to the left of $B_1$- and $D_2$-operators.
We note that the operator part of all the terms which appear from
this process can be expressed as
\begin{align}
\prod_{j \in S^{(2)}} B_0(u_j)
\prod_{j \in S^{(1)}} B_1(u_j) \prod_{j \in S^{(0)}} D_2(u_j),
\label{operatorpart}
\end{align}
for some tuple $(S^{(0)}, S^{(1)}, S^{(2)})$ of unordered sets
of integers $S^{(0)}$, $S^{(1)}$, $S^{(2)}$
satisfying \eqref{condition}.

We extract the coefficient of \eqref{operatorpart}
for a fixed tuple $(S^{(0)}, S^{(1)}, S^{(2)})$ as follows.
We first use \eqref{rtt3} and \eqref{rtt4} repeatedly and rewrite
the left hand side of \eqref{higherrankmultiplecommutationrelationsm=2}
as
\begin{align}
\prod_{j \in S^{(0)}} D_2(u_j) \prod_{j \in S^{(1)}} B_1(u_j) \prod_{j \in S^{(2)}} B_0(u_j).
\end{align}
We then use either \eqref{rtt1} or \eqref{rtt2} repeatedly 
and reverse the order of the types of operators.
In this process,
we always choose the first
term of the right hand side of \eqref{rtt1} and \eqref{rtt2}
when commuting the $B_0$-, $B_1$- and $D_2$-operators
to extract the coefficient
of \eqref{operatorpart}.
This is since if we once use the second term of the right hand side of \eqref{rtt1} or \eqref{rtt2}, we get other operators.
From this process, we find the coefficient of the operator is given by
$\displaystyle
\prod_{\substack{i \in S^{(0)} \\ j \in S^{(1)}}} \frac{1}{1-u_j/u_i}
\prod_{\substack{i \in S^{(0)} \\ j \in S^{(2)}}} \frac{1}{1-u_j/u_i}
\prod_{\substack{i \in S^{(1)} \\ j \in S^{(2)}}} \frac{1}{1-u_j/u_i}
$, hence \eqref{higherrankmultiplecommutationrelationsm=2} follows.

\end{proof}

\section{Yang-Baxter algebra and $K$-theoretic Gysin map
for partial flag bundles}

In this section,
combining the multiple commutation relations of the Yang-Baxter algebra
derived in the last section and description of the $K$-theoretic Gysin map
for flag bundles using symmetrizing operators,
we show the first and second type of partition functions on a rectangular grid,
which corresponds to Grothendieck classes of the Grothendieck group
of the partial flag bundles and a nonsingular variety respectively,
are directly connected via the $K$-theoretic Gysin map.
First we collect results from the geometry side.
There are various approaches to the Gysin map. See
\cite{AllPhD,All,AB,BV,CG,Nie,Gr,BFQ,BE,Prlecnote,PrBanach,
Tu,WebZie,Rim,AR1,AR2,RS,Zi1,Z2,Z3,Pr,DP1,DP2,HIMN,NN1,NN2,NN3}
as well as aforementioned articles for examples.

For a smooth scheme $X$, $K^0(X)$ denotes the Grothendieck group
of locally free coherent sheaves on $X$.
For $\mathcal{E}$ a locally free sheaf on $X$, we denote its class
in $K^0(X)$ by $[\mathcal{E}]$.

Let $E \longrightarrow X$ be a complex vector bundle
of rank $n$. We denote the bundles of flags of subspaces of dimensions
$q_1,\dots,q_m$ ($q_0:=0<q_1<q_2<\cdots< q_m<q_{m+1}:=n$) as
 $\pi_{q_1,\dots,q_m}:\mathcal{F\ell}_{q_1,\dots,q_m}(E) \longrightarrow X$. There exists a universal flag of subbundles of the pullback $\pi_{q_1,\dots,q_m}^*(E)$ of $E$ on $\mathcal{F\ell}_{q_1,\dots,q_m}(E)$,
\begin{align}
U_{q_0}:=0 \subsetneq U_{q_1} \subsetneq U_{q_2} \subsetneq \cdots
\subsetneq U_{q_m} \subsetneq U_{q_{m+1}}:=\pi_{q_1,\dots,q_m}^*(E),
\end{align}
where the rank of subbundle $U_{q_i}$ is $q_i$ for $i=0,1,\dots,m+1$.

The special case $m=n-1$, $q_j=j \ (j=1,2,\dots,n-1)$ of the flag bundle
$\pi_{1,2,\dots,n-1}:\mathcal{F\ell}_{1,2,\dots,n-1}(E) \longrightarrow X$
is called the complete flag bundle,
on which there exists the universal flag of subbundles
\begin{align}
0=U_0 \subsetneq U_1 \subsetneq \cdots \subsetneq U_{n-1} \subsetneq \pi_{1,2,\dots,n-1}^*(E).
\end{align}
We denote the class of the dual line bundle
$(U_i/U_{i-1})^\vee$ as $u_i$ in this section.

It is known that there are ways to describe Gysin map
using symmetrizing operators.
See \cite{All,BE,Prlecnote,PrBanach,NN1,NN3} for example.
One of the latest results is for the generalized cohomology theory by
Nakagawa-Naruse (\cite{NN1} Thm 4.10, \cite{NN3} Remark 3.9).
Specializing to the $K$-theory case is the following.
\begin{theorem}
Let $\pi_{q_1,\dots,q_m}$ be the partial flag bundle
$\pi_{q_1,\dots,q_m}: \mathcal{F \ell}_{q_1,\dots,q_m}(E) \longrightarrow X$.
The pushforward $\pi_{q_1,\dots,q_m *}:K^0(\mathcal{F \ell}_{q_1,\dots,q_m}
(E)) \longrightarrow K^0(X)$ of
a symmetric polynomial $f(1-t_1^{-1},\dots,1-t_n^{-1}) \in K^0(X)[t_1^{-1},\dots,t_n^{-1}]^{S_{q_1} \times S_{q_2-q_1} \times \cdots \times S_{n-q_m}}$ is given by
\begin{align}
&\pi_{q_1,\dots,q_m *} (f(1-u_1^{-1},\dots,1-u_n^{-1})) \nonumber \\
=&\sum_{\overline{w} \in S_n/S_{q_1} \times S_{q_2-q_1} \times \cdots \times S_{n-q_m}} w \cdot \Bigg[ 
\frac{f(1-u_1^{-1},\dots,1-u_n^{-1})}
{\prod_{k=1}^m \prod_{q_{k-1}<i \leq q_k} \prod_{q_{k}<j \leq n}(1-u_j/u_i)}
\Bigg]. \label{kgysinformulauvariables}
\end{align}
Here, $w$ in the right hand side of \eqref{kgysinformulauvariables}
acts as permutation
on the Grothendieck roots $(u_1,u_2,\dots,u_n)$,
and the sum is over all representatives of elements of
the fixed subgroup
$S_n/S_{q_1} \times S_{q_2-q_1} \times \cdots \times S_{n-q_m}$
of the symmetric group $S_n$.
\end{theorem}

We show the following formula for the $K$-theoretic Gysin map,
connecting two higher rank partition funtions
$F_{IJ}(u_1,\dots,u_n)$ \eqref{beforepushforwardpartitionfunction}
and $G_{IJ}(u_1,\dots,u_n)$ \eqref{afterpushforwardpartitionfunction}.
\begin{theorem} \label{maintheorem}
Let $\pi_{q_1,\dots,q_m}$ be the partial flag bundle
$\pi_{q_1,\dots,q_m}: \mathcal{F \ell}_{q_1,\dots,q_m}(E) \longrightarrow X$.
The pushforward $\pi_{q_1,\dots,q_m *}:K^0(\mathcal{F \ell}_{q_1,\dots,q_m}
(E)) \longrightarrow K^0(X)$ of
$F_{IJ}(u_1,\dots,u_n)$ is given by
\begin{align}
\pi_{q_1,\dots,q_m *} (F_{IJ}(u_1,\dots,u_n))
=G_{IJ}(u_1,\dots,u_n).
\label{mapofgrothendieckclassesflag}
\end{align}
\end{theorem}
\begin{proof}
We first note that $F_{IJ}(u_1,\dots,u_n)$ \eqref{beforepushforwardpartitionfunction} are polynomials in
$u_1^{-1},\dots,u_n^{-1}$ with integer coefficients.
This is since partition functions are constructed
from $R$-matrices, which means that in principle,
each $F_{IJ}(u_1,\dots,u_n)$
can be expressed as a sum of products of matrix elements of the $R$-matrices,
which are either $0, 1, u_j^{-1} \ (j=1,\dots,n)$ or $1-u_j^{-1} \ (j=1,\dots,n)$,
and expanding the expression, we note $F_{IJ}(u_1,\dots,u_n)$ 
are polynomials in $u_1^{-1},\dots,u_n^{-1}$.
Also note the symmetries stated in Lemma \ref{partialsymmetries}:
$F_{IJ}(u_1,\dots,u_n)$ are symmetric with respect to $u_{q_{j-1}+1},\dots,u_{q_j}$ for each $j=1,\dots,m+1$.

Applying the description of the $K$-theoretic Gysin map
using symmetrizing operators
\eqref{kgysinformulauvariables} to $F_{IJ}(u_1,\dots,u_n)$, we have
\begin{align}
&\pi_{q_1,\dots,q_m *} (F_{IJ}(u_1,\dots,u_n)) \nonumber \\
=&
\pi_{q_1,\dots,q_m *} ( \langle I|
\prod_{j=q_m+1}^{n} B_0(u_j)
\prod_{j=q_{m-1}+1}^{q_m} B_1(u_j)
\cdots
\prod_{j=q_{1}+1}^{q_2} B_{m-1}(u_j)
\prod_{j=1}^{q_1} D_m(u_j)
|J \rangle ) \nonumber \\
=&
\sum_{\overline{w} \in S_n/S_{q_1} \times S_{q_2-q_1} \times \cdots \times S_{n-q_m}} w \cdot \Bigg[ 
\frac{1}
{\prod_{k=1}^m \prod_{q_{k-1}<i \leq q_k} \prod_{q_{k}<j \leq n}(1-u_j/u_i)}
\nonumber \\
&\langle I|
\prod_{j=q_m+1}^{n} B_0(u_j)
\prod_{j=q_{m-1}+1}^{q_m} B_1(u_j)
\cdots
\prod_{j=q_{1}+1}^{q_2} B_{m-1}(u_j)
\prod_{j=1}^{q_1} D_m(u_j) |J \rangle
\Bigg]. \label{partfuncrep}
\end{align}
Since the (dual) basis vectors $|J \rangle$ and $\langle I|$
in \eqref{partfuncrep} are independent of $w$, we can move them out 
to get
\begin{align}
&\pi_{q_1,\dots,q_m *} (F_{IJ}(u_1,\dots,u_n)) \nonumber \\
=&
\langle I|
\sum_{\overline{w} \in S_n/S_{q_1} \times S_{q_2-q_1} \times \cdots \times S_{n-q_m}} w \cdot \Bigg[ 
\frac{1}
{\prod_{k=1}^m \prod_{q_{k-1}<i \leq q_k} \prod_{q_{k}<j \leq n}(1-u_j/u_i)}
\nonumber \\
&
\prod_{j=q_m+1}^{n} B_0(u_j)
\prod_{j=q_{m-1}+1}^{q_m} B_1(u_j)
\cdots
\prod_{j=q_{1}+1}^{q_2} B_{m-1}(u_j)
\prod_{j=1}^{q_1} D_m(u_j) \Bigg] |J \rangle.
\label{tochutoformula}
\end{align}
We then apply the key multiple commutation relations
of the Yang-Baxter algebra
\eqref{higherrankmultiplecommutationrelations}
to \eqref{tochutoformula} and get
\begin{align}
&\pi_{q_1,\dots,q_m *} (F_{IJ}(u_1,\dots,u_n)) \nonumber \\
=&
\langle I|
\prod_{j=1}^{q_1} D_m(u_j) \prod_{j=q_{1}+1}^{q_2} B_{m-1}(u_j)
\cdots \prod_{j=q_{m-1}+1}^{q_m} B_1(u_j) \prod_{j=q_m+1}^{n} B_0(u_j)
|J \rangle \nonumber \\
=&G_{IJ}(u_1,\dots,u_n).
\end{align}

\end{proof}

\begin{figure}[ht]
\includegraphics[width=8cm]{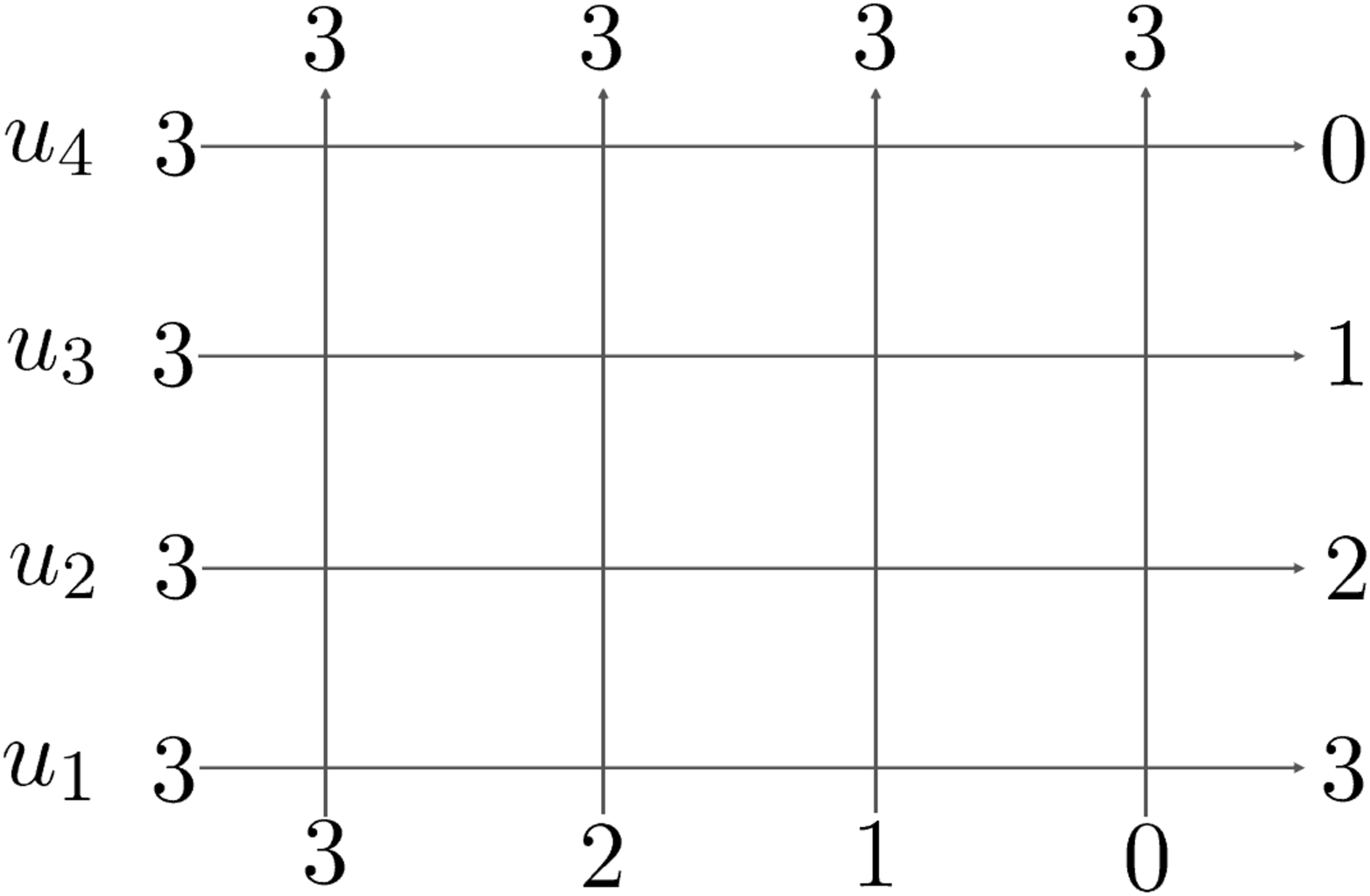}
\caption{The partition functions
$F_{IJ}(u_1,\dots,u_n)$ corresponding to
\eqref{completebeforepushforward}.
The figure represents the case $n=4$ ($m=3$).
}
\label{completeone}
\end{figure}

\begin{figure}[ht]
\includegraphics[width=8cm]{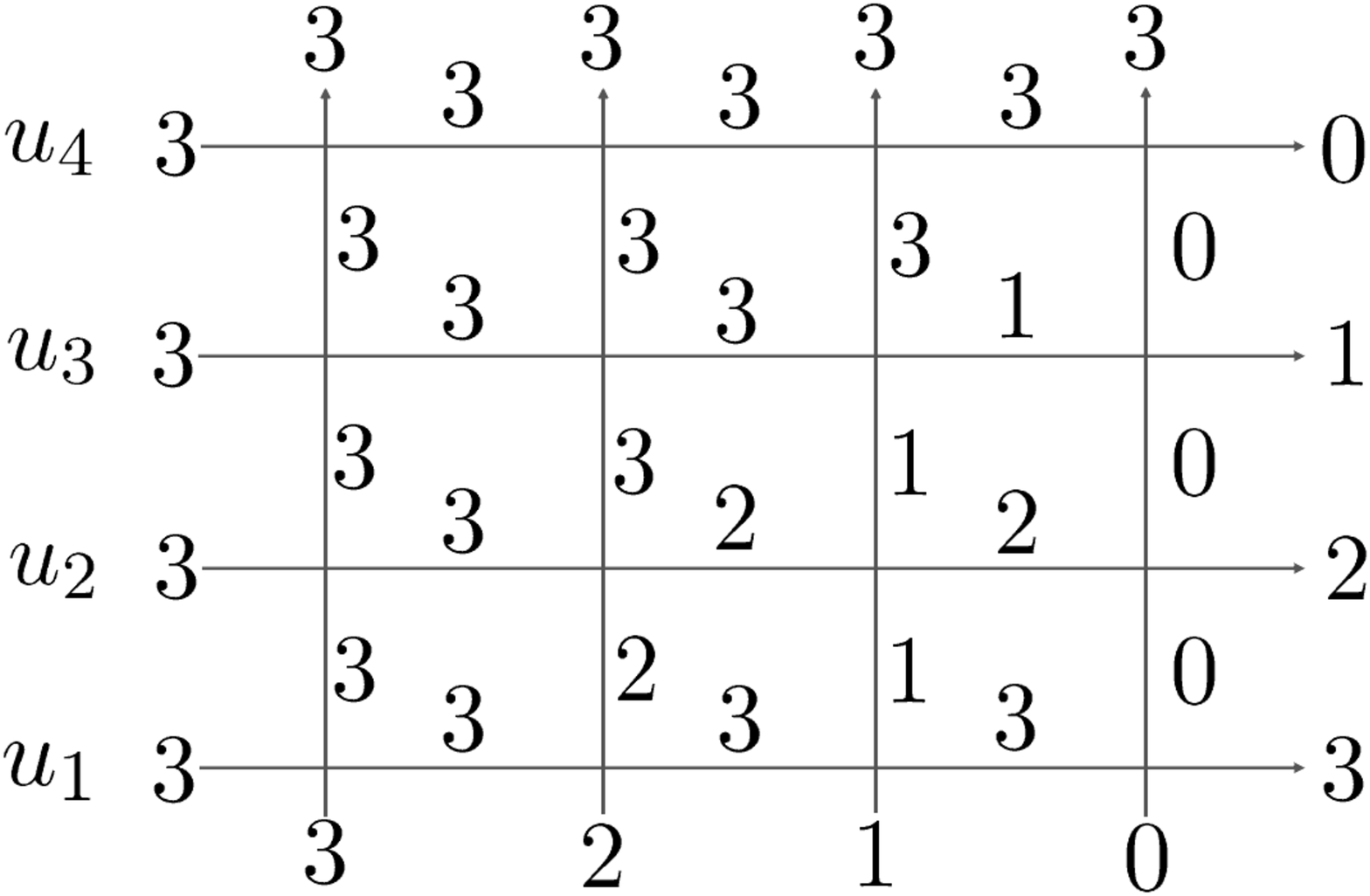}
\caption{The partition functions
$F_{IJ}(u_1,\dots,u_n)$ corresponding to
\eqref{completebeforepushforward}.
The figure represents the case $n=4$ ($m=3$).
Using Figure \ref{picturermatrix}, we find the product of the
$R$-matrix elements
in the first row is 1, the second row is $(1-u_3^{-1})$,
the third row is $(1-u_2^{-1})^2$, the fourth row is $(1-u_1^{-1})^3$,
and we find $F_{IJ}(u_1,u_2,u_3,u_4)=(1-u_1^{-1})^3 (1-u_2^{-1})^2 (1-u_3^{-1})$. For $n$ generic, we have  $F_{IJ}(u_1,\dots,u_n)=\prod_{j=1}^n (1-u_j^{-1})^{n-j}$.
}
\label{completetwo}
\end{figure}

The following is known for complete flag bundles (see \cite{NN1} for example)
\begin{align}
\displaystyle \pi_{1,2,\dots,n-1*}
\Bigg(
\prod_{j=1}^n (1-u_j^{-1})^{n-j}
\Bigg)
=1. \label{checkcomplete}
\end{align}

As a check of Theorem \ref{maintheorem}, let us show this.
This follows from the case $m=n-1$, $p=n$, $I=(n-1,n-1,\dots,n-1)$, $J=(n-1,n-2,\dots,0)$ of
Theorem \ref{maintheorem} as follows.
In this case, $F_{IJ}(u_1,\dots,u_n)$
\eqref{beforepushforwardpartitionfunction} is explicitly
\begin{align}
&F_{IJ}(u_1,\dots,u_n)=
{}_{1} \langle n-1| \otimes {}_{2} \langle n-1| \otimes \cdots \otimes {}_n \langle n-1| \nonumber \\
&\times 
B_0(u_n) B_1(u_{n-1}) \cdots B_{n-2}(u_2) D_{n-1}(u_1)|n-1 \rangle_1
\otimes |n-2 \rangle_2 \otimes \cdots \otimes |0 \rangle_{n},
\label{completebeforepushforward}
\end{align}
and
is graphically represented as Figure \ref{completeone}.
The figure represents the case $n=4$ ($m=3$).
In the same way with showing 
Lemma \ref{equivalencerelation} graphically,
one observes that only one configuration is allowed.
See Figure \ref{completetwo} for the case $n=4$ ($m=3$).
From Figure \ref{picturermatrix} and
Figure \ref{completetwo}, one notes that
the product of all the $R$-matrix elements appearing in
this unique configuration is given by $\prod_{j=1}^n (1-u_j^{-1})^{n-j}$,
i.e. we get 
\begin{align}
F_{IJ}(u_1,\dots,u_n)=\prod_{j=1}^n (1-u_j^{-1})^{n-j}.
\label{checkcompletebeforepushforward}
\end{align}

\begin{figure}[ht]
\includegraphics[width=8cm]{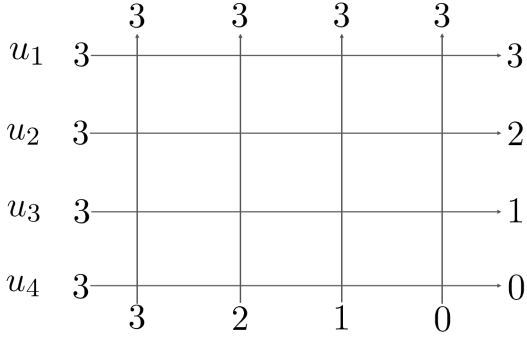}
\caption{The partition functions
$G_{IJ}(u_1,\dots,u_n)$ corresponding to
\eqref{completebeforepushforward}.
The figure represents the case $n=4$ ($m=3$).
}
\label{completethree}
\end{figure}

\begin{figure}[ht]
\includegraphics[width=8cm]{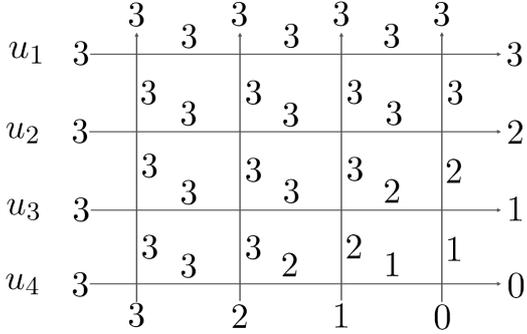}
\caption{The partition functions $G_{IJ}(u_1,\dots,u_n)$ corresponding to
\eqref{completebeforepushforward}.
The figure represents the case $n=4$ ($m=3$).
From Figure \ref{picturermatrix}, we find all the $R$-matrix elements
which appear in this unique configuration are 1, hence
we note $G_{IJ}(u_1,u_2,u_3,u_4)=1$. This also holds for 
$n$ generic, i.e. we have  $G_{IJ}(u_1,\dots,u_n)=1$.
}
\label{completefour}
\end{figure}

Next, we compute $G_{IJ}(u_1,\dots,u_n)$ for this case.
$G_{IJ}(u_1,\dots,u_n)$
\eqref{afterpushforwardpartitionfunction} is explicitly
\begin{align}
&G_{IJ}(u_1,\dots,u_n)=
{}_{1} \langle n-1| \otimes {}_{2} \langle n-1| \otimes \cdots \otimes {}_n \langle n-1| \nonumber \\
&\times 
D_{n-1}(u_1) B_{n-2}(u_2) \cdots B_1(u_{n-1})
B_0(u_n) |n-1 \rangle_1
\otimes |n-2 \rangle_2 \otimes \cdots \otimes |0 \rangle_{n},
\label{completeafterpushforward}
\end{align}
and
is graphically represented as Figure \ref{completethree}.
The figure represents the case $n=4$ ($m=3$).
One again notes that only one configuration is allowed, which is
Figure \ref{completefour}.
From Figure \ref{picturermatrix} and
Figure \ref{completefour}, we find
all the $R$-matrix elements appearing in
this unique configuration is 1, hence we conclude
\begin{align}
G_{IJ}(u_1,\dots,u_n)=1. \label{checkcompleteafterpushforward}
\end{align}
From \eqref{mapofgrothendieckclassesflag},
\eqref{checkcompletebeforepushforward} and
\eqref{checkcompleteafterpushforward}, we get \eqref{checkcomplete}.

\section{Inhomogenous version}
In this section, we discuss some generalization of
algebraic and geometric aspects discussed in the previous sections.
Let us introduce inhomogeneous parameters into the partition functions.
One can see from the Yang-Baxter equation \eqref{yangbaxter} that
the intertwining relation
\begin{align}
&R_{ab}(u_1,u_2) T_{a}(u_1|w_1,\dots,w_p) T_{b}(u_2|w_1,\dots,w_p)
\nonumber \\
=&T_{b}(u_2|w_1,\dots,w_p) T_{a}(u_1|w_1,\dots,w_p)R_{ab}(u_1,u_2),
\label{inhomogenousintertwiningrelation}
\end{align}
holds for the inhomogeneous version of the monodromy matrix
\eqref{monodromy}
\begin{align}
T_{a}(u|w_1,\dots,w_p)&=R_{a, p}(u,w_p) \cdots R_{a 1}(u,w_1),
\label{inhomogeneousmonodromy}
\end{align}
where parameters $w_1,\dots,w_p$ are introduced.
Recall the action of the $R$-matrix $R_{ij}(u,w)$
\eqref{rdefone}, \eqref{rdeftwo}, \eqref{rdefthree}.

The matrix elements of the inhomogeneous monodromy matrix
$T_{ij}(u|w_1,\dots,w_p):={}_a \langle j | T_{a}(u|w_1,\dots,w_p) | i \rangle_{a}$ satisfy the same commutation relations with the homogeneous version which follow from the intertwining relation.
For example, the relations
\eqref{rtt1}, \eqref{rtt2}, \eqref{rtt3}, \eqref{rtt4}, \eqref{rtt5}
with $D_m(u), B_k(u) \ (k=0,1,\dots,m-1)$ replaced by
$D_m(u|w_1,\dots,w_p):=T_{mm}(u|w_1,\dots,w_p), B_k(u|w_1,\dots,w_p):=T_{mk}(u|w_1,\dots,w_p) \ (k=0,1,\dots,m-1)$ hold.
Hence the inhomogeneous version of the multiple commutation relations
\eqref{higherrankmultiplecommutationrelations}
which are derived by those basic commutation relations
\begin{align}
&\displaystyle
\prod_{j=1}^{q_1} D_m(u_j|w_1,\dots,w_p) \prod_{j=q_{1}+1}^{q_2} B_{m-1}(u_j|w_1,\dots,w_p)
\cdots \nonumber \\
&\times
\prod_{j=q_{m-1}+1}^{q_m} B_1(u_j|w_1,\dots,w_p) \prod_{j=q_m+1}^{n} B_0(u_j|w_1,\dots,w_p)  \nonumber \\
=&\sum_{\overline{w} \in S_n/S_{q_1} \times S_{q_2-q_1} \times \cdots \times S_{n-q_m}} w \cdot \Bigg[ 
\frac{1}
{\prod_{k=1}^m \prod_{q_{k-1}<i \leq q_k} \prod_{q_{k}<j \leq n}(1-u_j/u_i)}
\nonumber \\
&\times 
\prod_{j=q_m+1}^{n} B_0(u_j|w_1,\dots,w_p)
\prod_{j=q_{m-1}+1}^{q_m} B_1(u_j|w_1,\dots,w_p)
\cdots \nonumber \\
&\times
\prod_{j=q_{1}+1}^{q_2} B_{m-1}(u_j|w_1,\dots,w_p)
\prod_{j=1}^{q_1} D_m(u_j|w_1,\dots,w_p)
\Bigg], \label{inhomogeneoushigherrankmultiplecommutationrelations}
\end{align}
hold as well.

We introduce the inhomogenous generalization of the first type of
partition functions
\eqref{beforepushforwardpartitionfunction}
\begin{align}
&F_{IJ}(u_1,\dots,u_n|w_1,\dots,w_p) \nonumber \\
=&
\langle I|
\prod_{j=q_m+1}^{n} B_0(u_j|w_1,\dots,w_p)
\prod_{j=q_{m-1}+1}^{q_m} B_1(u_j|w_1,\dots,w_p)
\cdots \nonumber \\
&\times \prod_{j=q_{1}+1}^{q_2} B_{m-1}(u_j|w_1,\dots,w_p)
\prod_{j=1}^{q_1} D_m(u_j|w_1,\dots,w_p) |J \rangle,
\label{inhomogeneousbeforepushforwardpartitionfunction}
\end{align}
and the second type of partition functions
\eqref{afterpushforwardpartitionfunction}
\begin{align}
&G_{IJ}(u_1,\dots,u_n|w_1,\dots,w_p) \nonumber \\
=&
\langle I|
\prod_{j=1}^{q_1} D_m(u_j|w_1,\dots,w_p) \prod_{j=q_{1}+1}^{q_2} B_{m-1}(u_j|w_1,\dots,w_p)
\cdots \nonumber \\
&\times \prod_{j=q_{m-1}+1}^{q_m} B_1(u_j|w_1,\dots,w_p) \prod_{j=q_m+1}^{n} B_0(u_j|w_1,\dots,w_p)
|J \rangle.
\label{inhomogenousafterpushforwardpartitionfunction}
\end{align}

\begin{figure}[ht]
\includegraphics[width=10cm]{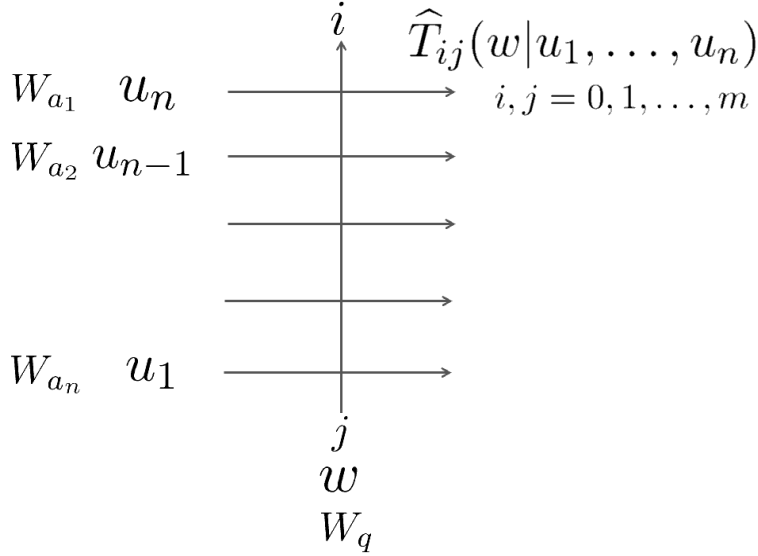}
\caption{The matrix elements
of the ``vertical" monodromy matrix
with respect to the quantum space $\widehat{T}_{ij}(w|u_1,\dots,u_n)$
\eqref{verticalmatrixelements}.
}
\label{pictureveticalmonodromy}
\end{figure}

At the algebraic level,
the following relation between
$F_{IJ}(u_1,\dots,u_n|w_1,\dots,w_p)$
and $G_{IJ}(u_1,\dots,u_n|w_1,\dots,w_p)$ follows
from
\eqref{inhomogeneoushigherrankmultiplecommutationrelations}
\begin{align}
&G_{IJ}(u_1,\dots,u_n|w_1,\dots,w_p) \nonumber \\
=&\sum_{\overline{w} \in S_n/S_{q_1} \times S_{q_2-q_1} \times \cdots \times S_{n-q_m}} w \cdot \Bigg[ 
\frac{F_{IJ}(u_1,\dots,u_n|w_1,\dots,w_p)}
{\prod_{k=1}^m \prod_{q_{k-1}<i \leq q_k} \prod_{q_{k}<j \leq n}(1-u_j/u_i)} \Bigg].
\end{align}
For $I$ and $J$ generic,
the inhomogeneous version of
partition functions $F_{IJ}(u_1,\dots,u_n|w_1,\dots,w_p)$
and $G_{IJ}(u_1,\dots,u_n|w_1,\dots,w_p)$ are not symmetric
with respect to variables $w_1,\dots,w_p$ in general.
Restricting to special cases, symmetries emerge
for smaller sets of variables.
For example,
from the standard argument used in the Yang-Baxter algebra,
one can see a symmetry emerges when 
the sequence of integers $i_k,i_{k+1},\dots,i_\ell$ in
$I=(i_1,\dots,i_{k},i_{k+1},\dots,i_\ell,\dots,i_p)$ ($1 \le k < \ell \le p$)
and $j_k,j_{k+1},\dots,j_\ell$ in
$J=(j_1,\dots,j_{k},j_{k+1},\dots,j_\ell,\dots,j_p)$ ($1 \le k < \ell \le p$)
are the same $i_k=i_{k+1}=\cdots=i_\ell=j_k=j_{k+1}=\cdots=j_\ell$.
In this case, the partition functions $F_{IJ}(u_1,\dots,u_n|w_1,\dots,w_p)$
and $G_{IJ}(u_1,\dots,u_n|w_1,\dots,w_p)$ are both symmetric with
respect to the variables $w_{k},w_{k+1},\dots,w_{\ell}$.
This can be shown using the Yang-Baxter algebra which is essentially the same
with the algebra appeared in the previous sections, but now we use
``vertical" monodromy matrix
\begin{align}
\widehat{T}(w|u_1,\dots,u_n)=R_{a_1 q}(u_n|w)R_{a_2 q}(u_{n-1}|w) \cdots
R_{a_n q}(u_1|w),
\end{align}
acting on $W_{a_1} \otimes W_{a_2} \otimes \cdots \otimes W_{a_n} \otimes W_q$.
The matrix elements of the ``vertical" monodromy matrix
with respect to the quantum space $W_q$ (Figure \ref{pictureveticalmonodromy})
\begin{align}
\widehat{T}_{ij}(w|u_1,\dots,u_n)={}_q \langle i|
\widehat{T}(w|u_1,\dots,u_n)
|j \rangle_q, \ \ \ i,j=0,1,\dots,m,
\label{verticalmatrixelements}
\end{align}
satisfy commutation relations
which are elements of the $RTT$ intertwining relation
in the vertical direction.
In particular, the commutativity
\begin{align}
[\widehat{T}_{ii}(w_1|u_1,\dots,u_n),
\widehat{T}_{ii}(w_2|u_1,\dots,u_n)
]=0, \ \ \ i=0,1,2,\dots,m, \label{verticalcommutativity}
\end{align}
follows from the intertwining relation.
Using the description of ``vertical" monodromy matrices,
one can see that if $i_k=i_{k+1}=\cdots=i_\ell=j_k=j_{k+1}=\cdots=j_\ell=:i$,
the $k \sim \ell$-th columns of the inhomogeneous version of
partition functions
$F_{IJ}(u_1,\dots,u_n|w_1,\dots,w_p)$
and $G_{IJ}(u_1,\dots,u_n|w_1,\dots,w_p)$
counted from left
are constructed from operators $\widehat{T}_{ii}(w_k|u_1,\dots,u_n)$,
$\widehat{T}_{ii}(w_{k+1}|u_1,\dots,u_n),\dots,
\widehat{T}_{ii}(w_\ell|u_1,\dots,u_n)$.
Thus, we note using the commutativity \eqref{verticalcommutativity} that
$F_{IJ}(u_1,\dots,u_n|w_1,\dots,w_p)$
and $G_{IJ}(u_1,\dots,u_n|w_1,\dots,w_p)$
are both symmetric with
respect to the variables $w_{k},w_{k+1},\dots,w_{\ell}$.

From the observation above, we have the following
formula for the $K$-theoretic Gysin map for example:
Let $E \longrightarrow X$ be a rank $n$ complex vector bundle
and $u_1,\dots,u_n$ be Grothendieck roots of $E^\vee$.
Let $H_j \longrightarrow X$ ($j=0,1,\dots,m$) be rank $h_j$ complex vector bundles
with Grothendieck roots $w_{\sum_{k=0}^{j-1} h_k+1},
w_{\sum_{k=0}^{j-1} h_k+2},
\dots,w_{\sum_{k=0}^j h_k}$ ($h_{-1}:=0$).
Take a positive integer $p$
satisfying $p \ge \sum_{k=0}^m h_k$.
Let $\pi_{q_1,\dots,q_m}$ be the partial flag bundle
$\pi_{q_1,\dots,q_m}: \mathcal{F \ell}_{q_1,\dots,q_m}(E) \longrightarrow X$
. For ordered sets of integers $I$ and $J$ which have the following forms
$I=(0^{h_0},1^{h_1},\dots,m^{h_m},i_{\sum_{k=0}^m h_k+1},\dots,i_p)$
and
$J=(0^{h_0},1^{h_1},\dots,m^{h_m},j_{\sum_{k=0}^m h_k+1},\dots,j_p)$,
the $K$-theoretic pushforward $\pi_{q_1,\dots,q_m *}:K^0(\mathcal{F \ell}_{q_1,\dots,q_m}
(E)) \longrightarrow K^0(X)$ of \\
$F_{IJ}(u_1,\dots,u_n|w_1,\dots,
w_{\sum_{k=0}^m h_k },1^{p-\sum_{k=0}^m h_k})$ is given by
\begin{align}
&\pi_{q_1,\dots,q_m *} (
F_{IJ}(u_1,\dots,u_n|w_1,\dots,
w_{\sum_{k=0}^m h_k },1^{p-\sum_{k=0}^m h_k})
) \nonumber \\
=&
G_{IJ}(u_1,\dots,u_n|w_1,\dots,
w_{\sum_{k=0}^m h_k },1^{p-\sum_{k=0}^m h_k}).
\label{inhomogeneousGysin}
\end{align}
The following special case of \eqref{inhomogeneousGysin}:
$m=1$, $h_1=0$, $J=(0^p)$ and
$I=(i_1,i_2,\dots,i_p)$ where $i_k=1$ for $k=\lambda_{n-q_1}+1,\lambda_{n-q_1-1}+2,\dots,\lambda_1+(n-q_1)$ $(h_0+1 \le \lambda_{n-q_1}+1 < \lambda_{n-q_1-1}+2< \cdots < \lambda_1+(n-q_1) \le p)$ and $i_k=0$ otherwise,
which corresponds to the case of Grassmann bundles, can be written down using
double Grassmannian Grothendieck polynomials as
\begin{align}
&\pi_{q_1 *}(G_{(\lambda_1,\dots,\lambda_{n-q_1})}
(1-u_{q_1+1}^{-1},\dots,1-u_n^{-1}|1-w_1,\dotsm1-w_{h_0},0,\dots)
\nonumber \\
&\times G_{p^{q_1}}(1-u_1^{-1},\dots,1-u_{q_1}^{-1}|1-w_1,\dots,1-w_{h_0},0,\dots)
) \nonumber \\
=&G_{((p-n+q_1)^{q_1},\lambda_1,\dots,\lambda_{n-q_1})}
(1-u_1^{-1},\dots,1-u_{n}^{-1}|1-w_1,\dots,1-w_{h_0},0,\dots).
\label{exampleinhomogeneousGysin}
\end{align}
Here, $G_\lambda(z_1,\dots,z_n|v_1,\dots)$ are
 the double Grassmannian Grothendieck polynomials
\cite{Buch1,LS,FK,Mc,IN,Buch2,Buch3} which have the following determinant form
\begin{align}
G_\lambda(z_1,\dots,z_n|v_1,\dots)=
   \frac{\mathrm{det}_n([z_i|\bs{v}]^{\lambda_j+n-j}(1-z_i)^{j-1})}
        {\prod_{1 \le i < j \le n}(z_i-z_j)},
\end{align}
where $\lambda=(\lambda_1,\lambda_2,\dots,\lambda_n)$
is a  partition, $\{z_1, \dots, z_n \}$ and $\bs{v}=\{v_1,\dots \}$ are sets of variables and
\begin{align}
[z_i|\bs{v}]^j=(z_i \oplus v_1) (z_i \oplus v_2)
\cdots (z_i \oplus v_j),
\end{align}
where $z \oplus v:=z+v- z v$.

One can show \eqref{exampleinhomogeneousGysin}
by slightly extending the argument given in \cite{MotegiGrassmann}
as follows.
First recall the following correspondence between
partition functions of the five-vertex models and Grothendieck polynomials.
We consider two types of partition functions
\begin{align}
\langle I|\prod_{j=1}^n B_0(u_j|w_1,\dots,w_p)|0^p \rangle,
\end{align}
and
\begin{align}
\langle I|\prod_{j=1}^m D_1(u_j|w_1,\dots,w_p) \prod_{j=1}^n B_0(u_{j+m}|w_1,\dots,w_p)|0^p \rangle,
\end{align}
where $I=(i_1,i_2,\dots,i_p)$ is $i_k=1$ for $k=\lambda_{n}+1,\lambda_{n-1}+2,\dots,\lambda_1+n$ $(1 \le \lambda_{n}+1 < \lambda_{n-1}+2< \cdots < \lambda_1+n \le p)$ and $i_k=0$ otherwise.
The first type of partition functions consisting of $B_0$-operators
are represented by Grothendieck polynomials
\cite{MS,MS2,GK,WZ,IN}
\begin{align}
&
\langle I|\prod_{j=1}^n B_0(u_j|w_1,\dots,w_p)|0^p \rangle \nonumber \\
=&G_{(\lambda_1,\dots,\lambda_n)}(1-u_1^{-1},\dots,1-u_n^{-1}|1-w_1,\dots,1-w_p,0,\dots).
\label{correspondencegrothendieckone}
\end{align}
Next, note that
Lemma \ref{equivalencetwotypes} can be extended to inhomogeneous
version as well, and we have the following equality
\begin{align}
&\langle I|\prod_{j=1}^m D_1(u_j|w_1,\dots,w_p) \prod_{j=1}^n B_0(u_{j+m}|w_1,\dots,w_p)|0^p \rangle \nonumber \\
&=
\langle \{ I,1^m \} |\prod_{j=1}^{m+n} B_0^{(p,m)}(u_j|w_1,\dots,w_p,1^m)|0^{p+m} \rangle, \label{auxcorrespondence}
\end{align}
where
\begin{align}
B_0^{(p,m)}(u|w_1,\dots,w_p,1^m)={}_a \langle 0|
R_{a,p+m}(u,1) \cdots R_{a,p+1}(u,1) R_{a,p}(u,w_p) \cdots R_{a1}(u,w_1)
|1 \rangle_a.
\end{align}
Combining \eqref{correspondencegrothendieckone} and \eqref{auxcorrespondence}
gives the following expression for the second type of partition functions
\begin{align}
&\langle I|\prod_{j=1}^m D_1(u_j|w_1,\dots,w_p) \prod_{j=1}^n B_0(u_{j+m}|w_1,\dots,w_p)|0^p \rangle \nonumber \\
&=G_{((p-n)^m,\lambda_1,\dots,\lambda_n)}(1-u_1^{-1},\dots,1-u_{m+n}^{-1}|1-w_1,\dots,1-w_p,0,\dots). \label{correspondencegrothendiecktwo}
\end{align}

We examine $F_{IJ}(u_1,u_2,\dots,u_n|w_1,\dots,w_{h_0},1^{p-h_0})$ and
$G_{IJ}(u_1,u_2,\dots,u_n|w_1,\dots,w_{h_0},1^{p-h_0})$ where
$J=(0^p)$.
First let us see
\begin{align}
&F_{IJ}(u_1,u_2,\dots,u_n|w_1,\dots,w_{h_0},1^{p-h_0}) \nonumber \\
=&\langle I|\prod_{j=q_1+1}^n B_0(u_j|w_1,\dots,w_{h_0},1^{p-h_0})
\prod_{j=1}^{q_1}D_1(u_j|w_1,\dots,w_{h_0},1^{p-h_0})|0^p \rangle.
\label{FIJspecialcase}
\end{align}

Using the action of the $D_1$-operators on $|0^p \rangle$
\begin{align}
D_1(u|w_1,\dots,w_p)|0^p \rangle
=\prod_{k=1}^p (1-w_k/u)|0^p \rangle,
\end{align}
we have
\begin{align}
&\prod_{j=1}^{q_1}D_1(u_j|w_1,\dots,w_{h_0},1^{p-h_0})|0^p \rangle
\nonumber \\
&=\prod_{j=1}^{q_1} \prod_{k=1}^{h_0} (1-w_k/u_j)
\prod_{j=1}^{q_1} \prod_{k=h_0+1}^{p} (1-1/u_j) |0^p \rangle \nonumber \\
&=G_{p^{q_1}}(1-u_1^{-1},\dots,1-u_{q_1}^{-1}|1-w_1,\dots,1-w_{h_0},0,\dots) |0^p \rangle,
\end{align}
\eqref{FIJspecialcase} becomes
\begin{align}
&F_{IJ}(u_1,u_2,\dots,u_n|w_1,\dots,w_{h_0},1^{p-h_0}) \nonumber \\
=&
G_{p^{q_1}}(1-u_1^{-1},\dots,1-u_{q_1}^{-1}|1-w_1,\dots,1-w_{h_0},0,\dots)
\langle I|\prod_{j=q_1+1}^n B_0(u_j|w_1,\dots,w_{h_0},1^{p-h_0})
|0^p \rangle. \label{FIJspecialcasetwo}
\end{align}
Applying the correspondence \eqref{correspondencegrothendieckone} to the right hand side
of \eqref{FIJspecialcasetwo}
(recall the sequence $I$ what we consider in \eqref{FIJspecialcasetwo}
is the following:
$I=(i_1,i_2,\dots,i_p)$ where $i_k=1$ for $k=\lambda_{n-q_1}+1,\lambda_{n-q_1-1}+2,\dots,\lambda_1+(n-q_1)$ $(h_0+1 \le \lambda_{n-q_1}+1 < \lambda_{n-q_1-1}+2< \cdots < \lambda_1+(n-q_1) \le p)$ and $i_k=0$ otherwise), we get
\begin{align}
&F_{IJ}(u_1,u_2,\dots,u_n|w_1,\dots,w_{h_0},1^{p-h_0}) \nonumber \\
=&
G_{(\lambda_1,\dots,\lambda_{n-q_1})}
(1-u_{q_1+1}^{-1},\dots,1-u_n^{-1}|1-w_1,\dotsm1-w_{h_0},0,\dots)
\nonumber \\
&\times G_{p^{q_1}}(1-u_1^{-1},\dots,1-u_{q_1}^{-1}|1-w_1,\dots,1-w_{h_0},0,\dots). \label{FIJspecialcaseGrothendieck}
\end{align}
Next we examine $G_{IJ}(u_1,u_2,\dots,u_n|w_1,\dots,w_{h_0},1^{p-h_0})$
\begin{align}
&G_{IJ}(u_1,u_2,\dots,u_n|w_1,\dots,w_{h_0},1^{p-h_0}) \nonumber \\
=&\langle I|
\prod_{j=1}^{q_1}D_1(u_j|w_1,\dots,w_{h_0},1^{p-h_0})
\prod_{j=q_1+1}^n B_0(u_j|w_1,\dots,w_{h_0},1^{p-h_0})|0^p \rangle.
\label{GIJspecialcase}
\end{align}
Applying the correspondence
\eqref{correspondencegrothendiecktwo} to the right hand side of
\eqref{GIJspecialcase} gives
\begin{align}
&G_{IJ}(u_1,u_2,\dots,u_n|w_1,\dots,w_{h_0},1^{p-h_0}) \nonumber \\
=&G_{((p-n+q_1)^{q_1},\lambda_1,\dots,\lambda_{n-q_1})}
(1-u_1^{-1},\dots,1-u_{n}^{-1}|1-w_1,\dots,1-w_{h_0},0,\dots).
\label{GIJspecialcaseGrothendieck}
\end{align}
\eqref{FIJspecialcaseGrothendieck} and \eqref{GIJspecialcaseGrothendieck}
gives \eqref{exampleinhomogeneousGysin}.

Finally we note
\eqref{exampleinhomogeneousGysin} is
also a special case of the following formula derived by Buch \cite{Buch1}.

\begin{theorem} (Buch \cite{Buch1}, Theorem 7.3) \label{theorembyBuch}
Let $\mathcal{A} \longrightarrow X$ and $\mathcal{B} \longrightarrow X$
be complex vector bundles of rank $n$ and $\ell$ respectively.
Let $\pi: \mathrm{Gr}_k(\mathcal{A}) \longrightarrow X$ be the Grassmann bundle of $k$-planes in $\mathcal{A}$ with universal tautological exact sequence $0 \rightarrow \mathcal{S} \rightarrow \pi^* \mathcal{A} \rightarrow \mathcal{Q} \rightarrow 0$ of vector bundles over $\mathrm{Gr}_k(\mathcal{A})$.
Let $\{\alpha_1,\dots,\alpha_n \}$,
$\{\beta_1,\dots,\beta_\ell \}$,
$\{ \sigma_1, \dots, \sigma_k \}$
and $\{ \omega_1,\dots,\omega_{n-k} \}$
be the sets of Grothendieck roots
for bundles $\mathcal{A}$, $\mathcal{B}$, $\mathcal{S}$ and
$\mathcal{Q}$ respectively.
Let $I=(I_1,\dots,I_{n-k})$ and $J=(J_1,J_2,\dots)$
be sequences of integers satisfying $I_j \geq \ell$ for all $j$.
Then we have
\begin{align}
\pi_*(G_I(\mathcal{Q}-\mathcal{B}) G_J(\mathcal{S}-\mathcal{B}))
=G_{I-(k)^{n-k},J}(\mathcal{A}-\mathcal{B}). \label{Buchformula}
\end{align}
\end{theorem}
Here, $G_I(\mathcal{A}-\mathcal{B})$ in \eqref{Buchformula}
are certain Grothendieck classes of $K^0(X)$
extended to sequences of integers \cite{Buch1}
from the case when $I$ are partitions
$I=\lambda=(\lambda_1,\dots,\lambda_i)$
satisfying $n \ge i$, in which case
$G_\lambda(\mathcal{A}-\mathcal{B})$
are the Grassmannian double Grothendieck polynomials
\begin{align}
G_\lambda(\mathcal{A}-\mathcal{B})
=G_\lambda(1-\alpha_1^{-1},\dots,1-\alpha_n^{-1}|1-\beta_1,\dots,1-\beta_\ell).
\end{align}

To compare \eqref{exampleinhomogeneousGysin}
with Theorem \ref{theorembyBuch}, let us first
recall that \eqref{exampleinhomogeneousGysin}
is the pushforward from the Grothendieck group
of $\mathcal{F\ell}_{q_1}(E)$.
There is an isomorphism 
$\mathrm{Gr}_{n-q_m}(E/U_{q_m}) \simeq \mathcal{F\ell}_{q_1,\dots,q_m}(E)$
between Grassmann bundles and partial flag bundles.
See \cite{NN3} Remark 3.3 for example.
We also use this to match with the former result for Grassmann bundles
\cite{MotegiGrassmann}.
Using this isomorphism,
we identify bundles $\mathcal{A}$, $\mathcal{S}$ and $\mathcal{Q}$
in Thm \ref{theorembyBuch} with bundles
$E^\vee$, $(E/U_{q_1})^\vee$ and $U_{q_1}^\vee$, respectively.
We also identify $\mathcal{B}$ with $H_0$.
Under these identifications, \eqref{exampleinhomogeneousGysin}
can be written as
\begin{align}
\pi_*(G_{p^{\mathrm{rank}(\mathcal{Q})}}
(\mathcal{Q}-\mathcal{B})
G_{\lambda}(\mathcal{S}-\mathcal{B}))=G_{((p-\mathrm{rank}(\mathcal{S}))^{\mathrm{rank}(\mathcal{Q})},\lambda)}(\mathcal{A}-\mathcal{B}),
\end{align}
where $p \ge h_0=\mathrm{rank}( \mathcal{B})$.
One can see this corresponds to the case $J=\lambda$, $I_1=\cdots=I_{n-k} \ge \ell$
of Theorem \ref{theorembyBuch}.

\section*{Acknowledgments}
This work was partially supported by grant-in-Aid
for Scientific Research (C) No. 21K03176 and No. 20K03793.


\begin{thebibliography}{00}
%
\bibitem{Bethe}
H. Bethe,
Z. Phys. {\bf 71}, 205 (1931).
%
\bibitem{LW}
E. H. Lieb and F.Y.Wu, Two-dimensional ferroelectric models, in Phase Transitions and Critical Phenomena (Academic
Press, London, 1972), Vol. 1, pp. 331-490.
%
\bibitem{Baxter}
R.J. Baxter,
{\it Exactly Solved Models in Statistical Mechanics}
(Academic Press, London, 1982).
%
\bibitem{KBI}
V.E. Korepin, N.M. Bogoliubov and A.G. Izergin,
{\it Quantum Inverse Scattering Method and Correlation functions}
(Cambridge University Press, Cambridge, 1993).
%
\bibitem{Dr}
V. Drinfeld,
Sov. Math. Dokl. {\bf 32}, 254 (1985).
%
\bibitem{J}
M. Jimbo,
Lett. Math. Phys. {\bf 10}, 63 (1985).
%
\bibitem{FST}
L. D. Faddeev, E. K. Sklyanin and L. A. Takhtajan,
Theor. Math. Phys. {\bf 40}, 194 (1979).
%
\bibitem{KRS}
P. P. Kulish, N. Yu. Reshetikhin and E. K. Sklyanin,
Lett. Math. Phys. {\bf 5}, 393 (1981).
%
\bibitem{GRTV}
V. Gorbounov, R. Rim\'anyi, V. Tarasov and A. Varchenko,
J. Geom. Phys. {\bf 74}, 56 (2013).
%
\bibitem{MS}
K. Motegi and K. Sakai,
J. Phys. A: Math. Theor. {\bf 46}, 355201 (2013).
%
\bibitem{MS2}
K. Motegi, K. and Sakai,
J. Phys. A: Math. Theor. {\bf 47}, 445202 (2014).
%
\bibitem{RTV}
R. Rim\'anyi, V. Tarasov and A. Varchenko,
J. Geom. Phys. {\bf 94}, 81 (2015).
%
\bibitem{AObethe}
M. Aganagic and A. Okounkov,
Moscow Math. J. {\bf 17}, 565 (2017).
%
\bibitem{Okounkov}
A. Okounkov,
Lectures on $K$-theoretic computations in enumerative geometry, In: Geometry of Moduli Spaces and Representation Theory, IAS/Park City Math. Ser., 24, Amer. Math. Soc., Providence, RI, (2017), pp. 251-380.
%
\bibitem{MO}
D. Maulik and A. Okounkov,
Quantum groups and quantum cohomology,
Ast\'erisque, {\bf 408} (2019).
%
\bibitem{AO}
M. Aganagic and A. Okounkov,
J. Amer. Math. Soc. {\bf 34}, 79 (2021).
%
\bibitem{GRV}
G. Felder, R. Rim\'anyi and A. Varchenko,
 SIGMA {\bf 14}, 132 (2018).
%
\bibitem{Konno1}
H. Konno,
J. Int. Syst. {\bf 2}, xyx011 (2017).
%
\bibitem{Konno2}
H. Konno,
J. Int. Syst. {\bf 3}, xyy012 (2018).
%
\bibitem{IIM}
T. Ikeda, S. Iwao and T. Maeno,
Int. Math. Res. Not. {\bf 19}, 6421 (2020).
%
\bibitem{PSZ}
P.P. Pushkar, A. Smirnov and A.M. Zeitlin,
Adv. Math. {\bf 360}, 106919 (2020).
%
\bibitem{KPSZ}
P. Koroteev, P.P. Pushkar, A. Smirnov and A.M. Zeitlin,
Sel. Math. New Ser. {\bf 27}, 87 (2021).
%
\bibitem{KZ}
P. Koroteev and A.M. Zeitlin,
Math.Res.Lett. {\bf 28}, 435 (2021).
%
\bibitem{SO}
A. Smirnov and A. Okounkov,
Invent. math. {\bf 229}, 1203 (2022).
%
\bibitem{KirillovSigma}
A.N. Kirillov,
SIGMA {\bf 12}, 034 (2016).
%
\bibitem{GK}
V. Gorbounov and C. Korff, 
Adv. Math. {\bf 313}, 282 (2017).
%
\bibitem{WZ}
M. Wheeler and P. Zinn-Justin,
J. Reine Angew. Math. {\bf 757}, 159 (2019).
%
\bibitem{BS}
V. Buciumas and T. Scrimshaw,
Int. Math. Res. Not. {\bf 10}, 7231 (2022).
%
\bibitem{BFHTW}
B. Brubaker, C. Frechette, A. Hardt, E. Tibor and K. Weber,
Frozen Pipes: Lattice Models for Grothendieck Polynomials,
arXiv:2007.04310.
%
\bibitem{ZJ}
P. Zinn-Justin,
SIGMA {\bf 14} paper 69, 48 (2018).
%
\bibitem{Yeliussizov}
D. Yeliussizov,
J. Comb. Th. A {\bf 161}, 453 (2019).
%
\bibitem{Iwao}
S. Iwao,
Alg. Comb. {\bf 3}, 1023 (2020).
%
\bibitem{Iwaoskew}
S. Iwao,
J. Alg. Comb. {\bf 56}, 493 (2022).
%
\bibitem{Iwaorefinedskew}
S. Iwao,
Free fermions and Schur expansions of multi-Schur functions,
arXiv:2105.02604.
%
\bibitem{MotegiGrassmann}
K. Motegi,
Nucl. Phys. B, {\bf 971}, 115513 (2021).
%
\bibitem{GMSZ}
W. Gu, L. Mihalcea, E. Sharpe and H. Zou
J. Geom. Phys. {\bf 177}, 104548 (2022).
%
\bibitem{Buch1}
A.S. Buch,
Duke Math. J. {\bf 115}, 75 (2002).
%
\bibitem{Pragacz}
P. Pragacz,
Ann. Sci. \'Ecole Norm. Sup. {\bf 21}, 413 (1988).
%
\bibitem{FP}
W. Fulton and P. Pragacz,
{\it Schubert varieties and degeneracy loci},
(Springer-Verlag, Berlin, 1998)
Appendix J by the authors in collaboration with
I. Ciocan-Fontanine.
%
\bibitem{LS}
A. Lascoux and M. Sch\"utzenberger,
Structure de {H}opf de
  l'anneau de cohomologie et de l'anneau de {G}rothendieck d'une vari\'et\'e de
  drapeaux,
C. R. Acad. Sci. Parix S\'er. I Math
295 (1982) 629.
%
\bibitem{FK}
S. Fomin and  A.N. Kirillov,
Grothendieck polynomials and the Yang-Baxter equation,
Proc. 6th Internat. Conf. on Formal Power Series and
Algebraic Combinatorics, DIMACS (1994) 183-190.
%
\bibitem{Mc}
P.J. McNamara,
Electron. J. Combin. {\bf 13}, 71 (2006).
%
\bibitem{IN}
T. Ikeda and H. Naruse,
Adv. Math. {\bf 243}, 22 (2013).
%
\bibitem{Buch2}
A.S. Buch,
Acta. Math. {\bf 189}, 37 (2002).
%
\bibitem{Buch3}
A.S. Buch,
Combinatorial $K$-theory, Topics in cohomological studies of algebraic varieties,
Trends Math., Birkh\"auser, Basel, 2005, pp. 87-103.
%
\bibitem{Buch4}
A.S. Buch,
Michigan Math. J. {\bf 57}, 93 (2008).
%
\bibitem{BuchMihalcea}
A.S. Buch, L.C. Mihalcea,
Duke Math. J. {\bf 156}, 501 (2011).
%
\bibitem{Lenart}
C. Lenart,
Ann. Comb. {\bf 4}, 67 (2000).
%
\bibitem{Ko}
V.E. Korepin,
Comm. Math. Phys. {\bf 86}, 391 (1982).
%
\bibitem{Iz}
A. Izergin,
Sov. phys. Dokl. {\bf 32}, 878 (1987).
%
\bibitem{Ku}
G. Kuperberg,
Int. Math. Res. Not. {\bf 3}, 139 (1996).
%
\bibitem{Ts}
O. Tsuchiya,
J. Math. Phys.
{\bf 39}, 5946 (1998).
%
\bibitem{PRS}
S. Pakuliak, V. Rubtsov and A. Silantyev,
J. Phys. A: Math. and Theor. {\bf 41}, 295204 (2008).
%
\bibitem{Ros}
H. Rosengren,
Adv. Appl. Math. {\bf 43}, 137 (2009).
%
\bibitem{Wheeler}
M. Wheeler,
Nucl. Phys. B {\bf 852}, 468 (2011).
%
\bibitem{Motegi}
K. Motegi,
J. Math. Phys. {\bf 59}, 053505 (2018).
%
\bibitem{HK1}
A. Hamel and R.C. King,
J. Alg. Comb.
{\bf 16}, 269 (2002).
%
\bibitem{Bogo}
N. M. Bogoliubov,
J. Phys. A: Math. and Gen. {\bf 38}, 9415 (2005).
%
\bibitem{ShigechiUchiyama}
K. Shigechi and M. Uchiyama,
J. Phys. A: Math. Gen. {\bf 38}, 10287 (2005).
%
\bibitem{BBF}
B. Brubaker, D. Bump and S. Friedberg,
Comm. Math. Phys. {\bf 308}, 281 (2011).
%
\bibitem{BMN}
D. Bump, P. McNamara and M. Nakasuji,
Comm. Math. Univ. Sancti Pauli. {\bf 63}, 23 (2014).
%
\bibitem{Lascoux}
A. Lascoux,
SIGMA {\bf 3}, 029 (2007).
%
\bibitem{Mcnamara}
P.J. McNamara,
Factorial Schur functions via the six-vertex model,
arXiv:0910.5288.
%
\bibitem{KS}
C. Korff and C. Stroppel,
Adv. Math. {\bf 225}, 200 (2010).
%
\bibitem{Korff}
C. Korff,
Lett. Math. Phys.
{\bf 104}, 771 (2014).
%
\bibitem{BW}
D. Betea and M. Wheeler,
J. Comb. Theory, Series A. {\bf 137}, 126 (2016).
%
\bibitem{BWZ}
D. Betea, M. Wheeler and P. Zinn-Justin,
J. Alg. Comb. {\bf 42}, 555 (2015).
%
\bibitem{WZ2}
M. Wheeler and  P. Zinn-Justin,
Adv. Math. {\bf 299}, 543 (2016).
%
\bibitem{Borodin}
A. Borodin,
Adv. Math. {\bf 306}, 973 (2017).
%
\bibitem{BP1}
A. Borodin and L. Petrov,
Sel. Math. New Ser. {\bf 24}, 751 (2018).
%
\bibitem{Takeyama}
Y. Takeyama, 
Funkcialaj Ekvacioj {\bf 61}, 349 (2018).
%
\bibitem{vDE}
J.F. van Diejen and E. Emsiz,
Comm. Math. Phys. {\bf 350}, 1017
(2017).
%
\bibitem{BBB}
B. Brubaker, V. Buciumas and D. Bump,
Communications in Number Theory and Physics.
{\bf 13}, 101 (2019).
%
\bibitem{BBBGduality}
B. Brubaker, V. Buciumas, D. Bump and N. Gray,
Appendix to \cite{BBB}.
%
\bibitem{BBBGdem}
B. Brubaker, V. Buciumas, D. Bump and H.P.A. Gustafsson,
J. Comb. Th. Series A, {\bf 178}, 105354 (2021).
%
\bibitem{BorodinWheeler}
A. Borodin and M. Wheeler,
Coloured stochastic vertex models and their spectral theory,
arXiv:1808.01866.
%
\bibitem{Zhong}
C. Zhong,
Lett. Math. Phys. {\bf 112}, 55 (2022).
%
\bibitem{FM}
O. Foda and M. Manabe,
J. High Energ. Phys. {\bf 2019}, 36 (2019).
%
\bibitem{Motegihigherrank}
K. Motegi,
J. Math. Phys. {\bf 61}, 053507 (2020).
%
\bibitem{ABW}
A. Aggarwal, A. Borodin and M. Wheeler,
Colored Fermionic Vertex Models and Symmetric Functions,
arXiv:2101.01605.
%
\bibitem{BBBG}
B. Brubaker, V. Buciumas, D. Bump and H.P.A. Gustafsson,
Metaplectic Iwahori Whittaker functions and supersymmetric lattice models,
arXiv:2012.15778.
%
\bibitem{GSFNR}
K. Motegi,
Nucl. Phys. B {\bf 954}, 114998 (2020).
%
\bibitem{FNR}
L.M. Feh\'er, A. N\'emethi and R. Rim\'anyi,
Comment. Math. Helv. {\bf 87}, 861 (2012).
%
\bibitem{GS}
P.L. Guo and S.C.C. Sun,
Adv. App. Math. {\bf 111}, 101933 (2019).
%
\bibitem{AllPhD}
J. Allman,
$K$-classes of quiver cycles, Grothendieck polynomials, and iterated
residues, PhD thesis, UNC-Chapel Hill, 2014,
%
\bibitem{All}
J. Allman,
Michigan Math. J. {\bf 63}, 865 (2014).
%
\bibitem{AB}
M. Atiyah and R. Bott,
Topology {\bf 23}, 1 (1984).
%
\bibitem{BV}
N. Berline and M. Vergne,
C. R. Acad. Sci. Paris, {\bf 295}, 539 (1982).
%
\bibitem{CG}
N. Chriss and V. Ginzburg,
{\it Representation Theory and Complex Geometry}.
Modern Birkh\"auser
Classics. Birkh\"auser Boston, 2009.
%
\bibitem{Nie}
H. A. Nielsen,
Bull. Soc. Math. France, {\bf 102}, 97 (1974).
%
\bibitem{Gr}
A. Grothendieck, Formule de Lefschetz, 
(R\'edig\'e par L. Illusie). In S\'eminaire de g\'eom\'etrie
alg\'ebrique du Bois-Marie 1965-66, SGA 5, Lect. Notes Math. 589,
Expos\'e No.III., pages 73-137.
1977.
%
\bibitem{BFQ}
P. Baum, W. Fulton, and G. Quart,
Acta Math. {\bf 143}, 193 (1979).
%
\bibitem{BE}
P. Bressler and S. Evens,
Trans. Amer. Math. Soc. {\bf 317}, 799 (1990).
%
\bibitem{Prlecnote}
P. Pragacz,
{\it Algebro-Geometric applications of Schur $S$- and $Q$-polynomials},
Topics in invariant theory
(Springer, Berlin, 1991)
(Paris, 1989/1990), 130-191, Lecture Notes in Math., {\bf 1478}.
\bibitem{PrBanach}
P. Pragacz,
{\it Symmetric polynomials and divided differences in formulas of
intersection theory}, Parameter Spaces (P. Pragacz, ed.),
{\bf 36}, Banach Center Publications, {\bf 125} (1996).
%
\bibitem{Tu}
L. Tu,
Computing the Gysin map using fixed points,
arXiv:1507.00283.
%
\bibitem{WebZie}
A. Weber and M. Zielenkiewicz,
J. Alg. Comb. {\bf 49}, 361 (2019).
%
\bibitem{Rim}
R. Rim\'anyi,
J. Alg. Comb. {\bf 40}, 527 (2014).
%
\bibitem{AR1}
J. Allman and R. Rim\'anyi, An iterated residue perspective on stable
Grothendieck polynomials, arXiv:1408.1911.
%
\bibitem{AR2}
J. Allman and R. Rim\'anyi,
$K$-theoretic Pieri rule via iterated residues,
S\'eminaire Lotharingien de Combinatoire 80B
(2018) Article 48.
%
\bibitem{RS}
R. Rim\'anyi and A. Szenes,
Residues, Grothendieck polynomials and $K$-theoretic Thom polynomials,
arXiv:1811.02055.
%
\bibitem{Zi1}
M. Zielenkiewicz, 
Centr. Eur. J. Math. {\bf 12}, 574 (2014).
%
\bibitem{Z2}
M. Zielenkiewicz,
J. Symp. Geom. {\bf 16}, 1455 (2018).
%
\bibitem{Z3}
M. Zielenkiewicz,
The Gysin homomorphism for homogeneous spaces via residues,
PhD Thesis, University of Warsaw, arXiv:1702.04123, 2017.
%
\bibitem{Pr}
P. Pragacz,
Proc. Amer. Math. Soc. {\bf 143}, 4705 (2015).
%
\bibitem{DP1}
L. Darondeau and P. Pragacz,
Int. J. Math. {\bf 28}, 1750077 (2017).
%
\bibitem{DP2}
L. Darondeau and P. Pragacz,
Fundamenta Mathematicae
{\bf 244}, 191 (2019).
%
\bibitem{HIMN}
T. Hudson, T. Ikeda, T. Matsumura and H. Naruse,
Adv. Math. {\bf 320}, 115 (2017).
%
\bibitem{NN1}
M. Nakagawa and H. Naruse,
Contemp. Math. {\bf 708}, 201 (2018).
%
\bibitem{NN2}
M. Nakagawa and H. Naruse,
Generating functions for the universal Hall-Littlewood 
$P$- and $Q$-functions, arXiv:1705.04791.
%
\bibitem{NN3}
M. Nakagawa and H. Naruse,
Math. Ann. {\bf 381}, 335 (2021).


\end{thebibliography}
\end{document}